# Balancing the trade-off between cost and reliability for wireless sensor networks: a multi-objective optimized deployment method

Long Chen · Yingying Xu · Fangyi Xu · Qian Hu · Zhenzhou Tang

**Abstract** The deployment of the sensor nodes (SNs) always plays a decisive role in the system performance of wireless sensor networks (WSNs). In this work, we propose an optimal deployment method for practical heterogeneous WSNs which gives a deep insight into the trade-off between the reliability and deployment cost. Specifically, this work aims to provide the optimal deployment of SNs to maximize the coverage degree and connection degree, and meanwhile minimize the overall deployment cost. In addition, this work fully considers the heterogeneity of SNs (i.e. differentiated sensing range and deployment cost) and three-dimensional (3-D) deployment scenarios. This is a multi-objective optimization problem, non-convex, multimodal and NP-hard. To solve it, we develop a novel swarm-based multi-objective optimization algorithm, known as the competitive multi-objective marine predators algorithm (CMOMPA) whose performance is verified by comprehensive comparative experiments with ten other state-of-the-art multi-objective optimization algorithms. The computational results demonstrate that CMOMPA is superior to others in terms of convergence and accuracy and shows excellent performance on multimodal multi-objective optimization problems. Sufficient simulations are also conducted to evaluate the effectiveness of the CMOMPA based optimal SNs deployment method. The results show that the optimized deployment can balance the trade-off among deployment cost, sensing reliability and network reliability. The source code is available on https://github.com/iNet-WZU/CMOMPA.

Corresponding author: Zhenzhou Tang.
Email: mr.tangzz@gmail.com

L. Chen, Y. Xu, F. Xu, Q. Hu and Z. Tang
College of Computer Science & Artificial Intelligence, Wenzhou University, Wenzhou, China, 325035.

## 1 Introduction

In recent years, wireless sensor networks (WSNs) have been widely used in industry, military, agriculture, health, and other fields [1], where the deployment of the sensor nodes (SNs) always plays a decisive role in the system performance. In practical WSNs, it is common that SNs may run out of power or become damaged and fail to work, which may lead to network disconnection or monitoring blind spots. Therefore, to ensure the effectiveness of sensing and the robustness of the network, an appropriate node deployment should be able to achieve the $K$-coverage over the target points to be monitored and the $C$-connectivity of the network.

Specifically, $K$-coverage means that each target point to be monitored is sensed by at least $K$ SNs so that $K - 1$ SNs failure can still guarantee the sensing coverage. On the other hand, $C$-connectivity means that each SN is wirelessly connected with other $C$ SNs to effectively avoid network bottleneck nodes and network segmentation. Fig. 1 illustrates some examples to more intuitively describe $K$-coverage and $C$-connectivity. SNs represented by black circles and blue circles are with different sensing ranges. $SN_1$ is 2-connectivity with $SN_2$, $SN_7$. $SN_5$ is 3-connectivity with $SN_4$, $SN_6$, and $SN_9$. Target point $t_1$ is 3-coverage with $SN_2$, $SN_7$, and $SN_8$. $t_2$ is 1-coverage with $SN_8$. $t_3$ realizes 0-coverage.

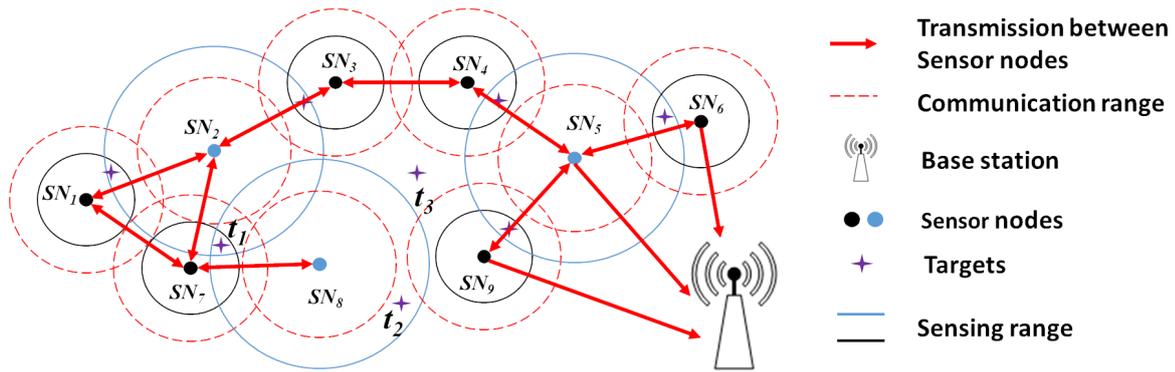

**Fig. 1** Examples of $K$-coverage and $C$-connectivity. $SN_1$ is 2-connectivity with $SN_2$, $SN_7$. $SN_5$ is 3-connectivity with $SN_4$, $SN_6$, and $SN_9$. Target point $t_1$ is 3-coverage with $SN_2$, $SN_7$, and $SN_8$. $t_2$ is 1-coverage with $SN_8$. $t_3$ realizes 0-coverage.

Studies on WSN deployment considering connectivity and coverage have been frequently reported. The review of literature on this area is summarized in Section 2 and Table 1. Although these existing works have done much research on optimizing the coverage and connectivity of WSNs, there are still some problems that remain open.

Firstly, they did not consider the deployment cost when optimizing the coverage and connectivity. For large-scale WSNs deployed in hard environments, such as forests and deep water, it is very important to carefully consider the total network costs, including node deployment cost and node manufacturing cost. Obviously, $K$-coverage, $C$-connectivity, and the total cost of network deployment are mutually restrictive goals. To be specific, in a given monitoring area, the more SNs deployed, the higher the connectivity and coverage, however, the higher the cost. Therefore, it is very meaningful to study the trade-off among $K$-coverage, $C$-connectivity, and the total cost. Secondly, the existing works ignored the fact that SNs in practical applications are typically heterogeneous. That is, there exist gaps in the sensing radius and deployment cost between different SNs. Thirdly, most works assumed that sensor nodes are deployed in a two-dimensional (2-D) environment, however, the deployment area in practical applications is often three-dimensional (3-D).

In view of the above problems, this paper investigates the optimal WSN deployment problem featuring the following highlights:

1) Three optimization objectives are considered in our work. Specifically, we aim to maximize the coverage degree which is defined as the number of SNs that can sense the target, the connection degree which is defined as the number of SNs that an SN can directly communicate with, and meanwhile minimize the overall deployment cost. Our work gives a deep insight into the trade-off among effectiveness, reliability and cost.
2) When modeling the network, we fully consider the heterogeneity of SNs, which is reflected in the differences of sensing ranges and deployment costs. Besides, the cost of deploying SNs also varies in different locations.
3) The deployment scenario is a 3-D environment, which is more in line with the actual deployment application.

The above optimization deployment problem is very complex, since it is a non-convex, multimodal multi-objective optimization problem. Besides, it is also a high-dimensional multi-objective optimization problem with large-scale decision variables because of the large number of SNs. It is an extremely tough challenge for classical convex optimization methods to find the Pareto optimal solutions. Fortunately, in recent years, swarm intelligence optimization algorithms such as the marine predators algorithm (MPA) [2], the artificial bee colony algorithm [3] and the particle swarm optimization algorithm [4], etc., have provided promising methods to search for the approximate optimal solutions of complex optimization problems [5, 6, 7, 8, 9, 10, 11, 12]. There have also been some multi-objective versions of swarm intelligence optimization algorithms developed to deal with multi-objective optimization problems, such as multi-objective artificial bee colony algorithm [13], multi-objective seagull optimization algorithm [14] and multi-objective whale optimization algorithm [15], etc. The effectiveness of these multi-objective swarm intelligence optimization algorithms has also been verified in scientific research and engineering practices [16, 17, 18, 19, 20, 21].

Multi-objective swarm intelligence optimization algorithms have two significant advantages for solving multi-objective optimization problems. Firstly, we can obtain a set of approximated Pareto optimal solutions, instead of one single in a single run. This set of solutions provides the entire range of solutions and the shape of the approximated Pareto-optimal front, which can more efficiently demonstrate the trade-off between the objectives to the decision-maker, who after that he/she chooses one out of the set of alternatives. Secondly, swarm intelligence optimization algorithms are not sensitive to the convexity of the objective functions and the hardness of the problems. It is particularly attractive to optimization problems in engineering since most of them are non-convex and NP-hard.

Given the above, we propose a multi-objective swarm intelligence optimization algorithm, known as CMOMPA (competitive multi-objective MPA), to solve the multi-objective optimization problem formulated in this work. The motivation behind using MPA is two-fold. On the one hand, balancing the trade-off between cost and reliability for a WSN with a large number of nodes is a multimodal optimization problem with considerably high-dimensional decision variables. MPA is very competitive in solving the above problems [2]. On the other hand, MPA has the advantages of low computational complexity, few parameters, and fast convergence, which has attracted many interests [22, 23, 24].

Moreover, CMOMPA incorporates the idea of a competitive swarm optimizer [25] to further improve its performance on multimodal multi-objective optimization problems. The competitive swarm optimizer updates the population by the competitive learning among individuals in each iteration, rather than the personal and/or global best individual. It has a better ability to exploit the small gap between two individuals whose fitness values are very close, which makes competitive swarm optimizer more powerful in dealing with multi-objective optimization problem with a large number of local optima [25]. In addition, CMOMPA uses the solutions selection method based on reference points [26], which can enhance the distribution diversity of solutions.

We have carried out extensive comparative experiments to verify the performance of CMOMPA on 20 popular benchmark functions selected from three popular benchmark suites on multi-objective optimization problems, i.e. ZDT [27], WFG [28], and DTLZ [29]. Moreover, CMOMPA has been compared with ten other state-of-the-art multi-objective optimization algorithms. The results show that the proposed CMOMPA outperforms other algorithms in terms of global convergence and accuracy.

To sum up, the main contributions of this paper are summarized as follows:

1) We build a practical optimal deployment model for WSNs. The model includes three optimization objectives, i.e., deployment cost, connection degree, and coverage degree. This optimization problem is constrained by the full coverage requirement of the monitoring area, the full connectivity of the network, and the minimum degree of coverage and connectivity. Moreover, the model fully considers the heterogeneity of SNs and 3-D deployment scenarios.

2) We propose a multi-objective swarm intelligence optimization algorithm, known as CMOMPA, to solve the above optimal deployment model. CMOMPA takes MPA as the optimizer and the reference point based method to determine elite individuals. In addition, CMOMPA leverages the individual competition mechanism to produce offspring with stronger diversity and distribution. CMOMPA can efficiently solve multimodal problems. The source code is available on https://github.com/iNet-WZU/CMOMPA.

3) Finally, we present comprehensive comparative experiments to evaluate the performance of the proposed CMOMPA. The results demonstrate that CMOMPA is superior to other algorithms in terms of convergence and accuracy and shows excellent performance for multimodal multi-objective optimization problems. In addition, we apply CMOMPA to the cost-saving optimal deployment scenario of WSNs. The results show that the optimized deployment can well balance the trade-off among deployment cost, sensing reliability, and network reliability.

For ease of reading, we have summarized the acronyms used in this paper and their full names as follows.

- **SNs**: Sensor nodes;
- **WSNs**: Wireless sensor networks;
- **MPA**: Marine predators algorithm
- **MOMPA**: Multi-objective MPA
- **CMOMPA**: Competitive multi-objective MPA
- **IGD**: Inverted generational distance
- **PFs**: Pareto fronts
- **NSGA-II**: Non-dominated sorting genetic algorithm II
- **NSGA-III**: Non-dominated sorting genetic algorithm III
- **PESA-II**: Pareto Envelope based Selection II
- **CMOPSO**: Competitive multi-objective particle swarm optimization
- **NSLS**: Non-dominated sorting and local search
- **CCMO**: Coevolutionary constrained multi-objective optimization
- **DCNSGA-III**: Dynamically constrained NSGA-III
- **MOEA/D**: Multi-objective evolutionary algorithm based on decomposition
- **CMOEA-MS**: Constrained multi-objective evolutionary algorithms with a two-stage

The rest of this paper is organized as follows. Section 2 introduces the related work on network deployment and multi-objective swarm intelligence optimization algorithms. Section 3 describes the system models and problem formulation. Section 4 presents the proposed CMOMPA and the optimal SN deployment method. The computational and simulation results and analysis are illustrated in Section 5. Finally, Section 6 concludes the paper and identifies the future work.

## 2 Related works

This section firstly reviews the existing models optimizing coverage and connectivity in WSNs. After that, we summarize some state-of-the-art multi-objective swarm intelligence optimization algorithms, particularly the existing multi-objective MPAs (MOMPAs), and highlight the necessity of proposing CMOMPA.

### 2.1 Optimal WSNs deployment

The optimal deployment of SNs has always been one of the hottest topics in the field of WSNs, where coverage and connectivity are two important optimization objectives.

In terms of coverage optimization, some works improved the coverage, network lifetime, energy consumption, and connectivity by optimally controlling the locations of mobile nodes [30, 31, 32, 37, 38]. On the other hand, heuristic algorithms have shown promising effectiveness in optimizing network coverage [33, 34, 35, 36, 39]. To maximize network coverage, a genetic algorithm based on Monte Carlo standard distribution method was used to deploy SNs in [33]. The study in [34] combined the greedy migration mechanism with the ant colony algorithm to deploy the SNs and improved the network coverage while reducing the deployment cost. The authors in [35] proposed a restrained Lloyd algorithm and a deterministic annealing algorithm to optimize the deployment of nodes in sensor networks and improve the overall performance of the network. In [36], a hybrid memetic framework was put forward to optimize the coverage of sensors so that the WSN can always maintain good coverage. In [39], a genetic algorithm was proposed to solve maximum coverage deployment in the 2-D environment.

In connectivity optimization, the authors in [42] proposed the path coloring-based connectivity detection algorithm and theoretically proves the superiority of the proposed algorithm. In [43], the connectivity of underwater optical sensor networks was analyzed using graph theory. In [40], the connectivity analysis models of WSNs under unicast and broadcast were provided. The study in [44] considered the deployment of sensor networks under the most general constraints. The study in [41] proposed three clustering algorithms to analyze the leading indicators of network connectivity, including availability indicators, time indicators, and reliability indicators. In [45], an improved non-dominated sorting genetic algorithm II (NSGA-II) algorithm was proposed to solve the connectivity problem of sensor networks to improve the network lifetime. A distributed connectivity restoration approach for a heterogeneous WSN was proposed in [46], which effectively improves the robustness of the network.

There have also been some works on the joint optimization of coverage and connectivity. In [53], the study took the coverage and connectivity of sensor networks into consideration comprehensively and proposed an optimal deployment method for a tunnel monitoring system. The study in [54] proposed a genetic algorithm based node placement method for 2-D WSNs, which improved the results by wavelets to achieve maximum coverage and connectivity. In [55], decentralized solutions were developed for determining and adjusting the levels of coverage and connectivity of a given network. In [56], an improved artificial bee colony algorithm was proposed to solve the optimal coverage and connectivity of WSNs. The [57] put forward the connectivity-based $k$-coverage hole detection algorithm, which can effectively monitor coverage holes and improve network coverage and connectivity. In [58], a sequential hybrid method was proposed to solve the coverage problem of sensor networks while taking into account connectivity. The study in [59] proposed four heuristic algorithms to solve the connectivity and coverage problems in wireless networks. In [60], an energy efficient wakeup scheduling scheme based on an improved Memetic algorithm was proposed, where four constraints, namely energy consumption, coverage, connectivity, and optimal length of wakeup schedule list were considered. The work in [61] ensured the coverage and connectivity of nodes in the network by miming a spider canvas.

Some works also consider the $K$-coverage of targets and the $C$-connectivity of nodes. In [47], the authors adopted the decomposition-based multi-objective algorithm to solve the problems of node $C$-connectivity and energy allocation in WSNs. In [48], considering different communication ranges of nodes, the authors proposed a $C$-connected scheme based on the injection of relay nodes in WSNs. The proposed method can better solve the problem of the connectivity of nodes. Both [47] and [48] considered the $C$-connectivity of nodes, but do not consider the $K$-coverage of targets. In [49], the authors proposed a novel algorithm called partial coverage learning automata,

**Table 1** Review of Literature: Coverage & connectivity optimizations.

| Ref | Objections | | | 2-D/3-D | Methodology |
|---|---|---|---|---|---|
| | Conn. | Cov. | Other. | | |
| [30] | | ✓ | | 2-D | Heuristic Algorithm |
| [31] | | ✓ | Network uniformity, deployment time | 2-D | Heuristic Algorithm |
| [32] | ✓ | ✓ | Energy consumption | 2-D | Mathematical optimization |
| [33] | | ✓ | Time to calculate the fitness function | 2-D | Heuristic Algorithm |
| [34] | | ✓ | Power consumption of SNs | 2-D | Heuristic Algorithm |
| [35] | | ✓ | | 2-D | Graph theory, Heuristic Algorithm |
| [36] | | ✓ | Coverage duration | 2-D | Heuristic Algorithm |
| [37] | | ✓ | Target detection | 2-D | Mathematical optimization |
| [38] | | ✓ | Target detection | 2-D | Mathematical optimization |
| [39] | | ✓ | | 2-D | Heuristic Algorithm |
| [40] | ✓ | | The scheme of broadcast,unicast | 3-D | Mathematical optimization |
| [41] | ✓ | | Indicator of availability,time,reliability | 2-D | Mathematical optimization |
| [42] | ✓ | | | 2-D | Path coloring, Mathematical optimization |
| [43] | ✓ | | | 2-D | Graph theory |
| [44] | ✓ | | Fixed budget | 2-D | Heuristic Algorithm |
| [45] | ✓ | ✓ | Residual energy of the SNs | 2-D | Heuristic Algorithm |
| [46] | ✓ | | Sent Bytes, movement cost of SNs | 2-D | Mathematical optimization |
| [47] | ✓ | | Power assignment | 2-D | Heuristic Algorithm |
| [48] | ✓ | | The number of relay nodes | 2-D | Graph theory |
| [49] | ✓ | ✓ | The number of sensors to activate | 2-D | Mathematical optimization |
| [50] | ✓ | ✓ | | 2-D | Heuristic Algorithm |
| [51] | ✓ | ✓ | | 2-D | Heuristic Algorithm |
| [52] | ✓ | ✓ | The number of SNs | 2-D | Heuristic Algorithm |
| [53] | ✓ | ✓ | | 2-D | Mathematical optimization |
| [54] | ✓ | ✓ | | 2-D | Heuristic Algorithm |
| [55] | ✓ | ✓ | | 2-D | Mathematical optimization |
| [56] | ✓ | ✓ | | 2-D | Heuristic Algorithm |
| [57] | ✓ | ✓ | | 2-D | Mathematical optimization |
| [58] | ✓ | ✓ | The number of SNs | 2-D | Graph theory, Heuristic Algorithm |
| [59] | ✓ | ✓ | | 2-D | Heuristic Algorithm |
| [60] | ✓ | ✓ | The wakeup scheduling scheme of SNs | 3-D | Heuristic Algorithm |
| [61] | ✓ | ✓ | | 3-D | Heuristic Algorithm |
| Ours | ✓ | ✓ | The deployment cost of network | 3-D | Heuristic Algorithm |

Symbol ✓ indicates that the corresponding objective has been optimized.
Conn. and Cov. are abbreviations for connectivity and coverage, respectively.

which ensures the energy of the network by adjusting the sleep state of nodes, to simultaneously maintain the network coverage and connectivity. However, the proposed method can only ensure effective coverage and connectivity within a given sub-area. In [50], the $K$-coverage and $C$-connectivity problem was handled by a mathematical model which integrated the Nelder–Mead method and the shuffled frog leading algorithm. In [51, 52], the authors had proposed solutions based on a single objective heuristic algorithm, considering the minimum number of nodes, $K$-coverage, and $C$-connectivity. In [50, 51, 52], the multi-objective problems were transformed into single objectives by weighted methods. But the weighted solution can not well reflect the trade-off between various objectives, and the heterogeneity of nodes was not considered. In [45], the authors dealt with four objective optimization problems by NSGA-II, i.e., minimizing the number of nodes, $K$-coverage, $C$-connectivity, and maximizing the energy level of nodes. However, the performance of the NSGA-II algorithm usually degrades when the number of optimization objectives is greater than two.

Table 1 summarizes the aforementioned literature. We can observe that none of them jointly optimize the cost, $K$-coverage, and $C$-connectivity. As far as we know, this work is the first to investigate the trade-off among the cost, network reliability, and sensing effectiveness.

2.2 Multi-objective swarm intelligence optimization algorithms

Recently, a lot of studies have been devoted to multi-objective swarm intelligence optimization algorithms. In general, a multi-objective swarm intelligence optimization algorithm involves two essential strategies, i.e., offspring generation strategy and elite selection strategy. Different algorithms may have different strategies. A multi-objective swarm intelligence optimization algorithm usually adopts the single-objective version of swarm intelligence optimization algorithm as the optimizer and generate offspring. The commonly used elite selection strategies are non-dominated sorting [69], the crowding distance approach [69], the neighborhood based method [70], the reference point based method [26], and so on.

**Table 2** Review of Literature: variants of multi-objective MPA.

| Ref | Offspring generation strategy | Elite selection strategy |
|---|---|---|
| [62] | MPA | Non-dominated sorting, Fuzzy set method |
| [63] | MPA, Gaussian Mutation | Non-dominated sorting, Crowded distance |
| [64] | MPA | Non-dominated sorting, Fuzzy set method |
| [65] | MPA, Non-uniform mutation operator | Non-dominated sorting, Niche approach |
| [66] | MPA | Non-dominated sorting, Niche approach |
| [67] | MPA | Non-dominated sorting, Crowded distance, |
| [68] | MPA, Gaussian perturbation | Non-dominated sorting, Reference points |
| Ours | MPA, Gaussian perturbation, Competition mechanism based learning strategy | Non-dominated sorting, Reference points |

In [13], the authors proposed a multi-objective version of artificial bee colony algorithm, which uses the artificial bee colony algorithm to generate offspring individuals and non-dominated sorting to select elite individuals. In [71], a multi-objective swarm intelligence optimization algorithm that combines differential-evolution and particle swarm optimization was proposed which uses a hybrid algorithm to to generate offspring individuals and non-dominated sorting to select elite individuals. A multi-objective seagull optimization algorithm was proposed in [14]. The authors used the seagull optimization algorithm to generate offspring individuals, and selected elite individuals using non-dominated sorting and grid methods to select elite individuals. In [72], the authors proposed multi-objective whale optimization algorithm, which uses the whale optimization algorithm to generate offspring individuals, and uses non-dominated sorting and Nelder-Mead simplex to select elite individuals. In [73], the authors proposed a multi-objective red deer algorithm, which uses a modified red deer algorithm to generate offspring individuals and selects elite individuals with roulette wheel selection. A hybrid multi-objective evolutionary algorithm was proposed in [74], which uses a crossover mutation operation to generate offspring individuals and selects elite individuals by comparing the Pareto front. In [75], the authors proposed a multi-objective optimization framework, which uses a hybrid swarm intelligence optimization algorithm to generate offspring individuals. In [76], the authors proposed a multi-objective scheduling fruit fly algorithm, which uses the fruit fly algorithm to generate offspring individuals. Both in [75] and [76], non-dominated sorting and the crowding distance approach were used for elite individuals selection.

There have also been some reports on multi-objective MPA, which have been summarized in Table 2. At the algorithm level, these algorithms are mainly different in the generation and selection strategy of offspring individuals. In [62], the authors proposed the two-objective MPA to deal with the problem of decreasing the emission of greenhouse gases and the fuel cost. In [65], an enhanced multi-objective optimization algorithm of MPA was proposed to handle problems with three objectives. In [64], the authors provided an improved multi-objective MPA for the optimized performance of combined alternating/direct current electrical grids. The multi-objective MPA proposed in [66] selects elite individuals by non-dominated sorting and a niche approach. In [63, 67], two versions of MOMPA based on crowding distance were put forward, however, its performance degrades when the number of objectives is more than two. In our previous work, we also proposed the MOMPA[68], which adopts the reference points based mechanism to select elite offsprings and achieves satisfactory performance in solution convergence and diversity for most multi-objective optimization problems, except for some complex multimodal problems.

According to the no free lunch theorem[77], there should be no any optimization algorithms that can be general-purpose universally-best. So, researchers are encouraged to develop more variety of optimization algorithms. There is a great demand for optimization algorithms to solve complex multi-objective optimization problems in practical engineering applications. These multi-objective optimization problems typically require optimization algorithms with excellent global and local search capabilities, uniform and complete solution distribution, and to be quite effective in dealing with multimodal objective functions. In view of the above, we developed CMOMPA, which uses a competitive mechanism based on learning strategies to generate offspring and utilizes a reference point-based strategy for individual selection. It has the attractive features of fast and accurate convergence and well distributed solutions.

## 3 System model and problem formulation

In this section, we firstly present the coverage model of a WSN, the communication model of an SN, and the overall deployment cost model. After that, we formulate the multi-objective optimization deployment problem based on the above three models.

### 3.1 Coverage model

Consider a deterministic deployment of a WSN. Let $\mathbf{L} = \{l_1, l_2, \cdots, l_i, \cdots, l_{\mathbb{I}_{\text{site}}}\}$ represent the set of the potential locations of SNs, where $\mathbb{I}_{\text{site}}$ is the total number of locations. In the 3-D target area $\mathcal{A}$, the coordinate of $l_i$ is denoted as $L_{l_i} = (x_{l_i}, y_{l_i}, z_{l_i})$. Let $\mathbf{T} = \{t_1, t_2, \cdots, t_j, \cdots, t_{\mathbb{I}_{\text{target}}}\}$ represent the set of target points to be sensed in $\mathcal{A}$, where $\mathbb{I}_{\text{target}}$ is the total target points. The area sensed by each SN of type $v$ is assumed to be a sphere with the radius of $R_s(v)$, as shown in Fig. 2.

Suppose that there are $\mathbb{I}_{\text{type}}$ types of SNs. We use $H_i^v$, $v \in \{1, 2, \cdots, \mathbb{I}_{\text{type}}\}$, $i \in \{1, 2, \cdots, \mathbb{I}_{\text{site}}\}$ to indicate whether the SN of type $v$ is deployed at the location $l_i$. If deployed, $H_i^v = 1$, otherwise it is 0. Moreover, only one SN can be placed at each potential location, that is, $\sum_{v=1}^{\mathbb{I}_{\text{type}}} H_i^v \leq 1$ for any given $l_i$. The deployment solution of a WSN can be denoted by

$$\mathbf{H} = [\mathcal{H}_1, \cdots, \mathcal{H}_i, \cdots, \mathcal{H}_{\mathbb{I}_{\text{site}}}], \tag{1}$$

where $\mathcal{H}_i = \left[H_i^1, H_i^2, \cdots, H_i^v, \cdots, H_i^{\mathbb{I}_{\text{type}}}\right]$.

We adopt the commonly used target point coverage model [51]. Specifically, let $s(l_i, t, v)$ be the boolean indicator indicating whether the target point $t$ can be effectively sensed by the SN of type $v$ located at $l_i$. $s(l_i, t, v)$ is determined as follows:

$$s(l_i, t, v) = \begin{cases} 1, & \|l_i, t\|_2 \leq R_s(v), \\ 0, & \text{others}. \end{cases} \tag{2}$$

We further define the joint sensing indicator indicating whether $t$ can be effectively sensed by at least one SN, denoted as $S(t)|\mathbf{L}$. Obviously,

$$S(t)|\mathbf{L} = 1 - \prod_{l_i \in \mathbf{L}} \left[1 - \sum_{v=1}^{\mathbb{I}_{\text{type}}} s(l_i, t, v) H_i^v\right]. \tag{3}$$

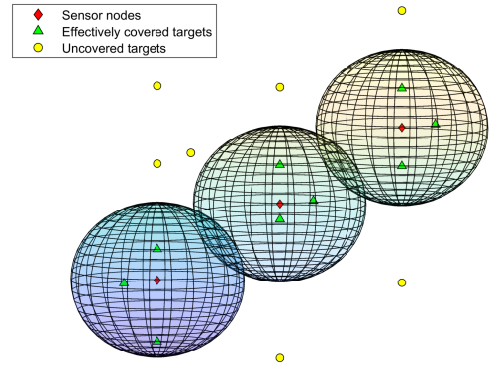

**Fig. 2** The spherical coverage model of sensor nodes. The yellow circles are uncovered targets because they exceed the sensed radiuses of all SNs.

Then, we can obtain the overall coverage rate of $\mathbf{T}$ as follows:

$$S_{\Sigma}|\mathbf{T} = \frac{1}{\mathbb{I}_{\text{target}}} \sum_{j=1}^{\mathbb{I}_{\text{target}}} S(t_j)|\mathbf{L}. \tag{4}$$

In addition, the total number of SNs that sense $t_j$ can be calculated as follows:

$$k_{\text{cov},t_j} = \sum_{l_i \in \mathbf{L}} \sum_{v=1}^{\mathbb{I}_{\text{type}}} s(l_i, t_j, v) H_i^v. \tag{5}$$

We take the average of $k_{\text{cov},t_j}$ over the target points set $\mathbf{T}$ as follows:

$$\sigma_{\text{cov}} = \frac{1}{\mathbb{I}_{\text{target}}} \sum_{j=1}^{\mathbb{I}_{\text{target}}} k_{\text{cov},t_j}. \tag{6}$$

### 3.2 Communication model

To be more in line with the practical applications, we adopt the probabilistic sensor communication model proposed in [78]. To be specific, let $\phi(l_i, l_j)_{i \neq j}$ denote the probability of effective communication between the two SNs located at $l_i$ and $l_j$:

$$\phi(l_i, l_j)_{i \neq j} = \begin{cases} 1, & \|l_i, l_j\|_2 \leq R_c - R_e, \\ e^{-\lambda_1 b^{\lambda_2}}, & |\|l_i, l_j\|_2 - R_c| < R_e, \\ 0, & \text{others}, \end{cases} \tag{7}$$

where $R_c$ denotes the communication range of the SN (We assume that all SNs have the same communication range), $R_e$ is a measure of the uncertainty in sensor detection, $b = \|l_i, l_j\|_2 - (R_c - R_e)$, and $\lambda_1$ and $\lambda_2$ are parameters that measure detection probability. For ease of understanding Eq. (7), Fig. 3 depicts

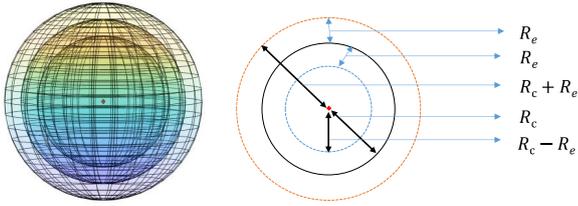

**Fig. 3** The left side is the probability communication model of an SN, and the right side is the cross-sectional graph of the communication model.

the probabilistic communication model and the cross-sectional graph.

If $\phi(l_i, l_j)_{i \neq j} \geq \delta$, where $\delta \in [0,1]$ is a given threshold, we say that the SN located at $l_i$ can communicate with the SN located at $l_j$. It can be described by the following Boolean model:

$$\omega(l_i, l_j) = \begin{cases} 1, & \phi(l_i, l_j)_{i \neq j} \geq \delta, \\ 0, & \text{others}. \end{cases} \quad (8)$$

The number of SNs that the SN located at $l_i$ can communicate with is

$$c_{\text{conn}, l_i} = \sum_{l_j \in \mathbf{L} \setminus l_i} \sum_{v=1}^{\mathbb{I}_{\text{type}}} \omega(l_i, l_j) H_i^v. \quad (9)$$

Thus, we obtain the average number of SNs that an SN can directly communicate with as follows:

$$\sigma_{\text{conn}} = \frac{1}{\sum_{i=1}^{\mathbb{I}_{\text{site}}} \sum_{v=1}^{\mathbb{I}_{\text{type}}} H_i^v} \sum_{i=1}^{\mathbb{I}_{\text{site}}} c_{\text{conn}, l_i}. \quad (10)$$

We assume that all wireless links are bi-directional so that the topology of the WSN can be abstracted as an undirected graph $G_{\text{WSN}}$. Let $\mathfrak{C}$ be the Boolean variable, that is, $\mathfrak{C}(G_{\text{WSN}}) = 1$ if $G_{\text{WSN}}$ is a connected graph, otherwise, $\mathfrak{C}(G_{\text{WSN}}) = 0$. $\mathfrak{C}$ can be obtained by the depth-first search method.

### 3.3 Cost model

We propose a total cost model for deploying a WSN in a 3-D heterogeneous environment, including node deployment cost, and node manufacturing cost. We use $C_{\text{SN}}(v), v = \{1, 2, \cdots, \mathbb{I}_{\text{type}}\}$ to represent the manufacturing cost of an SN of type $v$, and use $C_{\text{loc}}(i), i = \{1, 2, \cdots, \mathbb{I}_{\text{site}}\}$ to represent the deployment cost of each potential deployment location. The total deployment cost can be calculated as follows:

$$C_\Sigma = \sum_{i=1}^{\mathbb{I}_{\text{site}}} \sum_{v=1}^{\mathbb{I}_{\text{type}}} C_{\text{SN}}(v) \cdot C_{\text{loc}}(i) \cdot H_i^v. \quad (11)$$

### 3.4 Problem description

As mentioned above, in this work, we investigate the trade-off among cost, effectiveness and reliability by jointly optimizing three objectives, namely the overall cost, coverage degree ($K$-coverage) and connection degree ($C$-connectivity), under the constraints of complete coverage of the monitoring area and a connected network. To sum up, the cost-saving deployment problem of WSNs can be formulated as follows:

$$\begin{aligned}
& \min_{\mathbf{H}} \{C_\Sigma, -\sigma_{\text{cov}}, -\sigma_{\text{conn}}\}, \\
& \text{s.t. C1}: S_\Sigma | \mathbf{T} = 1, \\
& \quad\quad \text{C2}: \mathfrak{C}(G_{\text{WSN}}) = 1, \\
& \quad\quad \text{C3}: k_{\text{cov}, t_j} \geq K, \forall t_j, \\
& \quad\quad \text{C4}: c_{\text{conn}, l_i} \geq C, \forall i, \sum_{v=1}^{\mathbb{I}_{\text{type}}} H_i^v = 1 \\
& \quad\quad \text{C5}: \sum_{v=1}^{\mathbb{I}_{\text{type}}} H_i^v \leq 1, \forall\, i, \\
& \quad\quad \text{C6}: H_i^v \in \{0, 1\}, \forall\, i, v,
\end{aligned} \quad (12)$$

where $H_i^v$ is the decision vector defined in Eq. (1) whose search space is the set of vertices of a $\mathbb{I}_{\text{type}} \times \mathbb{I}_{\text{site}}$-dimensional unit hypercube, C1 ensures that all target points can be covered, C2 ensures the full network connectivity, C3 indicates that the coverage degree of any target point should be no less than a predefined value $K$, C4 indicates that the connection degree of any SN deployed in $l_i$ should be no less than a predefined value $C$, C5 ensures that only one SN can be placed at any potential location, and C6 indicates that $H_i^v$ must be a binary integer.

## 4 METHODOLOGY

Considering the multi-objective optimization problem formulated in Eq. (12) is a non-convex, discontinuous, multimodal and NP-hard problem, we develop the novel CMOMPA (competitive multi-objective MPA) to solve it. In this section, CMOMPA is first introduced in detail, including the initialization of the algorithm, iteration procedure, offspring generation strategy and complexity analysis. After that, we further present the CMOMPA-based optimal WSN deployment method.

### 4.1 CMOMPA

As the name suggests, CMOMPA is a multi-objective variant of the well-known single-objective optimizer MPA [2] that mimics the movement of marine predators and their prey using three phases. In our previous work, we have proposed a multi-objective MPA, known as MOMPA [68]. However, through extensive experiments, we found that the performance of MOMPA degrades in the face of multi-objective optimization problems with a large number of locally optimal solutions. Fig. 4 shows the Pareto fronts (PFs) of DTLZ1 and DTLZ3 benchmark functions obtained by MOMPA. We can see that they are far away from the true PFs, which indicates that the algorithm is trapped in the optimal local solution. In order to improve the global search ability, we propose an improved version of MOMPA, named CMOMPA, without increasing the complexity of the algorithm.

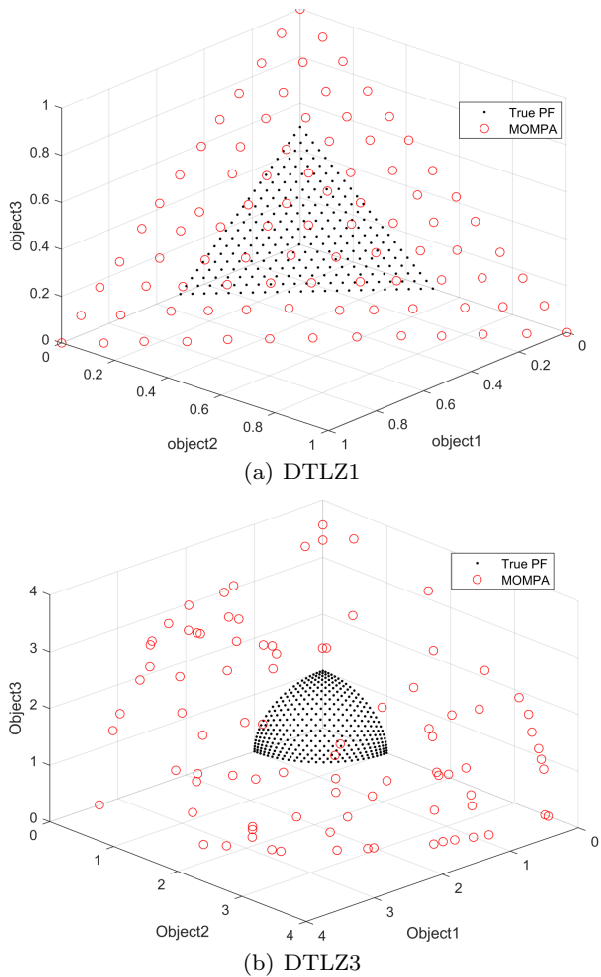

**Fig. 4** Comparison between the true PFs and the PFs obtained by MOMPA[68] in DTLZ1 and DTLZ3.

*4.1.1 Movement strategy*

CMOMPA involves two kinds of movements, namely the Lévy flight and Brownian motion. Lévy flight is a random walk in which the random step-lengths follow the Lévy distribution as follows:

$$R_L = \frac{0.05c}{|b|^{1/\lambda}}, \quad (13)$$

where $R_L$ is the random step-length; $c$ and $b$ are normally distributed, $c \sim N(0, \sigma_c{}^2)$, $b \sim N(0, 1)$, $\lambda = 1.5$, and $\sigma_c$ is given by

$$\sigma_c = \left[\frac{\Gamma(1+\lambda) \cdot \sin(\frac{\pi\lambda}{2})}{\Gamma(\frac{1+\lambda}{2}) \cdot \lambda \cdot 2^{\frac{(\lambda-1)}{2}}}\right]^{1/\lambda}, \quad (14)$$

where $\Gamma(x) = \int_0^\infty t^{x-1}e^{-x}\mathrm{d}t$ represents the Gamma function. Brownian motion is a Gaussian process with mean zero and variance one.

*4.1.2 Initialization*

In CMOMPA, we use $\mathbf{P_0}$ to denote the initial population, which is described as follows:

$$\mathbf{P_0} = [\mathbf{X}_1, \mathbf{X}_2, \cdots, \mathbf{X}_N]^T, \quad (15)$$

where $\mathbf{X}_i = [X_{i,1}, X_{i,2}, \cdots, X_{i,d}]$ is the $i$th individual in $\mathbf{P}_0$, $d$ represents the dimension of the problem, and $N$ is the number of individuals in the population. Each individual represents a feasible solution. The initial solution $\mathbf{X}_i$ is uniformly distributed over the search space as follows:

$$\mathbf{X}_i = \mathbf{X}_{\min} + \mathrm{rand}() \cdot (\mathbf{X}_{\max} - \mathbf{X}_{\min}), \quad (16)$$

where $\mathbf{X}_{\min}$ and $\mathbf{X}_{\max}$ are the lower and upper bounds of the solution space, and rand() returns a random number in $(0, 1)$.

After that, we select $N_a$ solutions in $\mathbf{P}_0$ by the reference points based method to create the initial archive $\mathbf{A}_0$ which is defined as follows:

**Definition 1** Archive ($\mathbf{A}$). The archive $\mathbf{A}$ is a set that is used to store the $N_a$ best solutions obtained in each iteration. The size of $\mathbf{A}$ equals the number of reference points, which can be calculated as

$$N_a = \binom{m+p-1}{p}, \quad (17)$$

where $m$ is the number of optimization objectives, and $p$ is the number of divisions on each objective in the normalized hyperplane.

Then, we obtain the set of non-dominated individuals of $\mathbf{A}_0$ by the non-dominated sorting method, which is used to construct the initial predator matrix ($\mathbf{E}_1$) whose size is $N_a$ by Algorithm 1 ($k$=0).

**Algorithm 1** Update $\mathbf{E}_k$ in the $k$th iteration.

**Input:** The set of non-dominated individuals of $\mathbf{A}_0$: $\mathbf{A}_k^{\text{nd}}$
**Output:** $\mathbf{E}_k$
1: $\mathbf{E}_k = \{ \underbrace{\mathbf{A}_k^{\text{nd}}, \ldots, \mathbf{A}_k^{\text{nd}}}_{\text{Repeat } \left\lfloor \frac{N_a}{\|\mathbf{A}_k^{\text{nd}}\|_2} \right\rfloor \text{ times}} \}$.
2: **if** $N_a$ is not an integer multiple of $\|\mathbf{A}_k^{\text{nd}}\|_2$ **then**
3:     Randomly select $\{N_a - \left\lfloor \frac{N_a}{\|\mathbf{A}_k^{\text{nd}}\|_2} \right\rfloor \|\mathbf{A}_k^{\text{nd}}\|_2\}$ individuals from $\mathbf{A}_k^{\text{nd}}$.
4:     Add these non-dominated individuals to $\mathbf{E}_k$.
5: **return** $\mathbf{E}_k$

**Algorithm 2** Gaussian elite perturbation

**Input:** $\mathbf{P}_k$
**Output:** $\mathbf{P}^{\text{GEP}}$
1: **for all** individuals in $\mathbf{P}_k$ **do**
2:     Randomly select a dimension $j, j = 1, 2, \ldots, d$.
3:     Generate the new individual for $\mathbf{P}_k$ by Eq. (22).
4: **return** $\mathbf{P}^{\text{GEP}}$

#### 4.1.3 The iterative process

In CMOMPA, the iterative process is divided into three stages. Specifically, the first stage is from the first iteration to the $(k_{\max}/3)$th iteration, where $k_{\max}$ is the maximum number of iterations. For the $k$th $(1 \le k < k_{\max}/3)$ iteration, the moving step matrix of the preys, denoted as $\mathbf{S}_k$, and the updated $\mathbf{P}_k$ can be obtained as follows:

$$\begin{aligned} \mathbf{S}_k &= \mathbf{R}_B \odot (\mathbf{E}_k - \mathbf{R}_B \odot \mathbf{A}_{k-1}), \\ \mathbf{P}_k &= \mathbf{A}_{k-1} + \theta \cdot \mathbf{R} \odot \mathbf{S}_k, \end{aligned} \quad (18)$$

where $\theta$ is a constant, $\mathbf{R}$ is a random vector in which each element is given by rand(), $\mathbf{R}_B$ is the Brownian motion vector and $\odot$ is the dot product operator.

The second stage is from the $(k_{\max}/3)$th iteration to the $(2k_{\max}/3)$th iteration. In this stage, $\mathbf{S}_k$ and $\mathbf{P}_k$ $(k_{\max}/3 \le k \le 2k_{\max}/3)$ are updated as follows.

$$\begin{aligned} \mathbf{S}_k \Big|_1^{\lfloor \frac{N_a}{2} \rfloor} &= \mathbf{R}_L \odot \left( \mathbf{E}_k \Big|_1^{\lfloor \frac{N_a}{2} \rfloor} - \mathbf{R}_L \odot \mathbf{A}_{k-1} \Big|_1^{\lfloor \frac{N_a}{2} \rfloor} \right), \\ \mathbf{S}_k \Big|_{\lfloor \frac{N_a}{2} \rfloor + 1}^{N_a} &= \mathbf{R}_B \odot \left( \mathbf{R}_B \odot \mathbf{E}_k \Big|_{\lfloor \frac{N_a}{2} \rfloor + 1}^{N_a} - \mathbf{A}_{k-1} \Big|_{\lfloor \frac{N_a}{2} \rfloor + 1}^{N_a} \right), \\ \mathbf{P}_k \Big|_1^{\lfloor \frac{N_a}{2} \rfloor} &= \mathbf{A}_{k-1} \Big|_1^{\lfloor \frac{N_a}{2} \rfloor} + \theta \cdot \mathbf{R} \odot \mathbf{S}_k \Big|_1^{\lfloor \frac{N_a}{2} \rfloor}, \\ \mathbf{P}_k \Big|_{\lfloor \frac{N_a}{2} \rfloor + 1}^{N_a} &= \mathbf{E}_k \Big|_{\lfloor \frac{N_a}{2} \rfloor + 1}^{N_a} + \theta \cdot \gamma \odot \mathbf{S}_k \Big|_{\lfloor \frac{N_a}{2} \rfloor + 1}^{N_a}, \end{aligned} \quad (19)$$

where $\mathbf{R}_L$ is the random step length vector in Lévy flight pattern; $\mathbf{R}_B$ is the random step length vector in Brownian movement pattern; $\lfloor \cdot \rfloor$ is the floor operator; $\mathbf{V}\big|_a^b$ returns the sub-vector of $\mathbf{V}$ from the $a$th element to the $b$th element, and $\gamma$ is an adaptive parameter changing with the increase of $k$ as follows:

$$\gamma = \left(1 - \frac{k}{k_{\max}}\right)^{\frac{2k}{k_{\max}}}. \quad (20)$$

The last stage is from the $(2k_{\max}/3)$th iteration to the end. In this stage, $\mathbf{S}_k$ and $\mathbf{P}_k$ $(k > 2k_{\max}/3)$ can be obtained as follows:

$$\begin{aligned} \mathbf{S}_k &= \mathbf{R}_L \odot (\mathbf{R}_L \odot \mathbf{E}_k - \mathbf{A}_{k-1}), \\ \mathbf{P}_k &= \mathbf{E}_k + \theta \cdot \gamma \odot \mathbf{S}_k. \end{aligned} \quad (21)$$

#### 4.1.4 Gaussian elite perturbation

In the $k$th iteration, the Gaussian elite perturbation strategy is applied to $\mathbf{P}_k$ to generate another $N_a$ individuals. We denote this new individual set as $\mathbf{P}^{\text{GEP}}$. For the $i$th individual in $\mathbf{P}_k$, select a dimension randomly, namely the $j$th dimension, then recalculate the value of $X_{i,j}$ as follows.

$$X_{i,j}^{\text{GEP}} = X_{i,j} + (X_{\max,j} - X_{\min,j}) G, \quad (22)$$

where $G \sim N(0,1)$, $X_{\max,j}$ and $X_{\min,j}$ are the upper and lower bounds of the $j$-dimensional variables, respectively. The pseudo-code of the Gaussian elite perturbation is presented in Algorithm 2.

#### 4.1.5 Competition mechanism based learning strategy

CMOMPA leverages the idea of the competition mechanism based learning strategy [25] to improve global search ability. The competition mechanism based learning consists of two components, namely pairwise competition and individual learning. Specifically, in the $k$th iteration, CMOMPA randomly selects one individual each from $\mathbf{P}_k$ and $\mathbf{P}^{\text{GEP}}$, respectively, and calculates their Fitness values by the shift-based density estimation [79] strategy as follows.

$$\text{Fitness}(\mathbf{X}) = \min_{\mathbf{Y} \in \mathbf{P} \setminus \{\mathbf{X}\}} \sqrt{\sum_{i=1}^{m} (\max\{0, f_i(\mathbf{Y}) - f_i(\mathbf{X})\})^2}, \quad (23)$$

where $\mathbf{X}$ is an individual in the population $\mathbf{P}$, $f_i(\mathbf{X})$ denotes the $i$th objective value of $\mathbf{X}$ and $m$ denotes the number of objectives. The shift-based density estimation strategy can evaluate the quality of a solution in terms of convergence and diversity.

After that, the fitness values of these two individuals are compared. The individual with a worse fitness value is a loser and is recorded as $\mathbf{X}^l$. It is updated according

**Algorithm 3** Competition mechanism based learning strategy

**Input:** $\mathbf{P}$, $\mathbf{P}^{\text{GEP}}$.
**Output:** $\mathbf{P}^{\text{CML}}$
1: Fitness ← Calculate the Fitness of each individual in $\mathbf{P}$ and $\mathbf{P}^{\text{GEP}}$ by Eq.(23).
2: $\mathbf{P}^{\text{CML}} \leftarrow \emptyset$.
3: **while** $\mathbf{P} \neq \emptyset$ and $\mathbf{P}^{\text{GEP}} \neq \emptyset$ **do**
4:     Randomly select an individual $\mathbf{p}$ from $\mathbf{P}$.
5:     Randomly select an individual $\mathbf{q}$ from $\mathbf{P}^{\text{GEP}}$.
6:     $\mathbf{P} \leftarrow \mathbf{P}\backslash\mathbf{p}$, $\mathbf{P}^{\text{GEP}} \leftarrow \mathbf{P}^{\text{GEP}}\backslash\mathbf{q}$.
7:     **if** Fitness ($\mathbf{p}$) < Fitness ($\mathbf{q}$) **then**
8:         $\mathbf{X}^l \leftarrow \mathbf{p}$.
9:         $\mathbf{X}^w \leftarrow \mathbf{q}$.
10:    **else**
11:        $\mathbf{X}^l \leftarrow \mathbf{q}$.
12:        $\mathbf{X}^w \leftarrow \mathbf{p}$.
13:    Update $\mathbf{X}^l$ by learning from $\mathbf{X}^w$ by Eq. (24).
14:    Mutate $\mathbf{X}^l$ and $\mathbf{X}^w$ by polynomial mutation.
15:    $\mathbf{P}^{\text{CML}} \leftarrow \mathbf{P}^{\text{CML}} \cup \{\mathbf{X}^l, \mathbf{X}^w\}$
16: **return** $\mathbf{P}^{\text{CML}}$

**Algorithm 4** Elite selection method

**Input:** $\mathbf{Q}$
**Output:** $\mathbf{A}$
1: $l \leftarrow 1$.
2: **while** $|\mathbf{Q}| > 0$ **do**
3:     **for all** $\mathbf{X}_i \in \mathbf{Q}$ **do**
4:         **if** $\nexists \mathbf{X}_j \succ \mathbf{X}_i (j \neq i)$ **then**
5:             $\mathcal{L}_l \leftarrow \mathcal{L}_l \cup \mathbf{X}_i$.
6:     $\mathbf{Q} \leftarrow \mathbf{Q} - \mathcal{L}_l$.
7:     $l \leftarrow l + 1$.
8: $\mathbf{A} \leftarrow \emptyset, l \leftarrow 1$.
9: **while** $|\mathbf{A} \cup \mathcal{L}_l| \leq N_a$ **do**
10:    $\mathbf{A} \leftarrow \mathbf{A} \cup \mathcal{L}_l$.
11:    $l \leftarrow l + 1$.
12: **if** $|\mathbf{A}| < N_a$ **then**
13:    Sort all individuals in $\mathcal{L}_l$ in descending order by reference point based method.
14:    $\mathbf{A} \leftarrow \mathbf{A} \cup \mathcal{L}_l(1 : N_a - |\mathbf{A}|)$.
15: **return** $\mathbf{A}$.

to the individual with a better fitness, denoted as $\mathbf{X}^w$, as follows.

$$\mathbf{X}^l = \mathbf{X}^l + \eta\left(\mathbf{X}^w - \mathbf{X}^l\right), \qquad (24)$$

where $\eta$ is a uniformly random value in $[0, 1]$. Besides, to further improve the performance of CMOMPA in exploration and exploration capability, both $\mathbf{X}^l$ and the updated $\mathbf{X}^w$ are slightly mutated by polynomial mutation [80].

This procedure repeats until all individuals in $\mathbf{P}_k$ and $\mathbf{P}^{\text{GEP}}$ are updated. Algorithm 3 depicts the procedure of the competition mechanism based learning strategy.

*4.1.6 Elite selection*

**Definition 2** File ($\mathbf{Q}$).
The file $\mathbf{Q}$ is a set that is used to store all individuals generated in each iteration. Specifically, in the $k$th iteration, $\mathbf{Q}_k = \mathbf{A}_{k-1} \cup \mathbf{P}_k \cup \mathbf{P}^{\text{GEP}}_k \cup \mathbf{P}^{\text{CML}}_k$. Individuals in $\mathbf{Q}$ are candidates for elite selection.

**Definition 3** Dominate.
An individual $\mathbf{X}$ is said to dominate another one $\mathbf{X}'$ if both of the following conditions are true, 1) $\mathbf{X}$ is no worse than $\mathbf{X}'$ in all objectives, and 2) $\mathbf{X}$ is strictly better than $\mathbf{X}'$ in at least one objective. $\mathbf{X}$ dominating $\mathbf{X}'$ is denoted as $\mathbf{X} \succ \mathbf{X}'$.

In the $k$th iteration, we obtain $\mathbf{Q}_k$ containing $5N_a$ individuals. After that, we need to select $N_a$ individuals among $\mathbf{Q}_k$ to generate $\mathbf{A}_k$. Firstly, we use the non-dominated sorting method to divide $\mathbf{Q}_k$ into several levels, namely $\mathcal{L}_1, \mathcal{L}_2, \ldots$. For two different levels, namely $\mathcal{L}_i$ and $\mathcal{L}_j (i < j)$, any individual in $\mathcal{L}_i$ dominates all individuals in $\mathcal{L}_j$, whereas individuals in the same level do not dominate each other. Then, all the individuals in $\bigcup_{i=1}^{l} \mathcal{L}_i (l \geq 1)$ are added to $\mathbf{A}_k$ where $l$ satisfies the following conditions: 1) $\left|\bigcup_{i=1}^{l} \mathcal{L}_i\right| \leq N_a$; and 2) $\left|\bigcup_{i=1}^{l+1} \mathcal{L}_i\right| > N_a$. If $\left|\bigcup_{i=1}^{l} \mathcal{L}_i\right|$ equals to $N_a$, then $\mathbf{A}_k = \bigcup_{i=1}^{l} \mathcal{L}_i$. Otherwise, the first $N_a - \left|\bigcup_{i=1}^{l} \mathcal{L}_i\right|$ best individuals in $\mathcal{L}_{l+1}$ are further added to $\mathbf{A}_k$. In this case, the reference point-based method proposed in [26] is employed to choose elite individuals in $\mathcal{L}_{l+1}$. The pseudo-code of elite selection is presented in Algorithm 4.

*4.1.7 CMOMPA and computational complexity*

The overall procedure of CMOMPA is described as follows.

**Initialization:**

Step1: Initialize $\mathbf{P}_0$ by Eq. (16), $\mathbf{Q}_1 \leftarrow \emptyset$. Select $N_a$ solutions in $\mathbf{P}_0$ by the reference points based method to create $\mathbf{A}_0$.

Step2: Construct the initial predator matrix $\mathbf{E}_1$ by Algorithm 1.

**Iterations**($k \geq 1$):

Step3: In the $k$th iteration, generate $\mathbf{P}_k$ by Eq. (18), (19), or (21), accordingly.

Step4: Generate $\mathbf{P}^{\text{GEP}}_k$ by Algorithm 2.

Step5: Generate $\mathbf{P}^{\text{CML}}_k$ by Algorithm 3.

Step6: Generate $\mathbf{Q}_k = \mathbf{A}_{k-1} \cup \mathbf{P}_k \cup \mathbf{P}^{\text{GEP}}_k \cup \mathbf{P}^{\text{CML}}_k$.

Step7: Generate $\mathbf{A}_k$ by Algorithm 4, $k = k + 1$.

Step8: Update $\mathbf{E}_k$ by Algorithm 1.

Step9: Repeat the loop from Step 3 to 8 until the maximum number of iterations is reached.

**Algorithm 5** CMOMPA procedure
1: $k \leftarrow 0$, initialize $\mathbf{P}_k$ by Eq. (16).
2: Select $N_a$ solutions in $\mathbf{P}_k$ to construct $\mathbf{A}_k$ by the reference points based method.
3: Initialize $\mathbf{E}_1$ by Algorithm 1.
4: **for** $k = 1$ to $k_{\max}$ **do**
5:    **if** $k < k_{\max}/3$ **then**
6:       Update $\mathbf{P}_k$ by Eq. (18).
7:    **else if** $k \leq 2k_{\max}/3$ **then**
8:       Update $\mathbf{P}_k$ by Eq. (19).
9:    **else**
10:      Update $\mathbf{P}_k$ by Eq. (21).
11:    Generate $\mathbf{P}_k^{\text{GEP}}$ by Algorithm 2.
12:    Generate $\mathbf{P}_k^{\text{CML}}$ by Algorithm 3.
13:    $\mathbf{Q}_k \leftarrow \mathbf{A}_{k-1} \cup \mathbf{P}_k \cup \mathbf{P}_k^{\text{GEP}} \cup \mathbf{P}_k^{\text{CML}}$.
14:    Generate $\mathbf{A}_k$ by Algorithm 4, $k = k+1$.
15:    Update $\mathbf{E}_k$ by Algorithm 1.
16: **return** Approximate Pareto optimal solutions $\mathbf{A}_k$.

Algorithm 5 describes the detailed algorithm of CMOMPA.

The complexity of CMOMPA is within our acceptable range. We use $n$ to represent the size of the population, $d$ to represent the dimension of the decision variable, and $m$ to represent the number of objectives. The three-stage evolutionary complexity of the population is $O(n \times d)$. Gaussian disturbance complexity is $O(n)$. The complexity of non-dominated sorting and solution selection based on reference point is $O(m \times n^2)$. The complexity of the competition mechanism based learning strategy is $O(n)$. The complexity of the polynomial mutation is $O(n)$. We use $cof_1$ to represent the computational complexity of the benchmark function, so the computational complexity of CMOMPA is $O(k_{max}(n \times d + n + cof_1 \times n + m \times n^2))$.

### 4.2 CMOMPA based optimal deployment

#### 4.2.1 Application model

The CMOMPA has been discussed in detail above. Next, the optimal deployment method based on CMOMPA is presented in this subsection.

Considering $H_i^v \in \{0,1\}$, $i \in \{1, 2, \ldots, \mathbb{I}_{\text{site}}\}$, $v \in \{1, 2, \ldots, \mathbb{I}_{\text{type}}\}$, is a binary integer. To solve the solution of Eq. (12), $H_i^v$ is relaxed as a real number $\tilde{H}_i^v \in [0,1]$. Concretely, $H_i^v \to \tilde{H}_i^v \in [0,1]$ with the constraint of $\tilde{H}_i^v - \left(\tilde{H}_i^v\right)^2 \leq \epsilon$, where $\epsilon$ is a sufficiently small number. Consequently, the search space of $\tilde{H}_i^v$ is also extended to the entire $\mathbb{I}_{\text{type}} \times \mathbb{I}_{\text{site}}$-dimensional unit hypercube and the problem formulated in Eq. (12) can be transformed as follows:

$$\min_{\widetilde{\mathbf{H}}} \{C_\Sigma, -\sigma_{\text{cov}}, -\sigma_{\text{conn}}\},$$

$$\begin{aligned}
\text{s.t. } & C1 : S_\Sigma \mid \mathbf{T} = 1, \\
& C2 : \mathfrak{C}(G_{\text{WSN}}) = 1, \\
& C3 : k_{\text{cov}, t_j} \geq K, \forall\, t_j, \\
& C4 : c_{\text{conn}, l_i} \geq C, \forall\, i, round\left(\sum_{v=1}^{\mathbb{I}_{\text{type}}} \tilde{H}_i^v\right) = 1 \\
& C5 : \sum_{v=1}^{\mathbb{I}_{\text{type}}} \tilde{H}_i^v \leq 1, \forall\, i, \\
& C6 : \tilde{H}_i^v \in [0,1], \forall\, i, v, \\
& C7 : \tilde{H}_i^v - \left(\tilde{H}_i^v\right)^2 \leq \epsilon, \forall\, i, v,
\end{aligned} \quad (25)$$

where $\widetilde{\mathbf{H}} = \left\{\tilde{\mathcal{H}}_1, \cdots, \tilde{\mathcal{H}}_i, \cdots, \tilde{\mathcal{H}}_{\mathbb{I}_{\text{site}}}\right\}$ is the relaxed $\mathbf{H}$, $round(\cdot)$ is a rounding operation. It is worth noting that the main difference between Eq. (12) and Eq. (25) is that Eq. (25) relaxes the discrete variables $H_i^v$ into continuous variables $\tilde{H}_i^v$ subject to C6 of C7.

#### 4.2.2 Constraint processing

The problem optimal deployment problem formulated in Eq. (25) is a multi-objective optimization problem with constraints. Consequently, the constrained-domination should be used to judge the dominant relationship between solutions during elite individual selection presented in Algorithm 4, instead of the unconstrained domination. The constrained-domination is defined as follows:

**Definition 4** Feasible and Infeasible solution.
An solution $\widetilde{\mathbf{H}} = \left\{\tilde{\mathcal{H}}_1, \cdots, \tilde{\mathcal{H}}_i, \cdots, \tilde{\mathcal{H}}_{\mathbb{I}_{\text{site}}}\right\}$ is said to be a feasible solution of the optimal deployment problem formulated in Eq. (25) if it satisfies the constraints C1 - C7. Otherwise, it is an infeasible solution.

**Definition 5** Constrained-Dominate.
A solution $\widetilde{\mathbf{H}} = \left\{\tilde{\mathcal{H}}_1, \cdots, \tilde{\mathcal{H}}_i, \cdots, \tilde{\mathcal{H}}_{\mathbb{I}_{\text{site}}}\right\}$ is said to constrained-dominate another solution $\widetilde{\mathbf{H}}' = \left\{\tilde{\mathcal{H}}'_1, \cdots, \tilde{\mathcal{H}}'_i, \cdots, \tilde{\mathcal{H}}'_{\mathbb{I}_{\text{site}}}\right\}$ if any of the following conditions is true, 1) both are feasible solutions and $\widetilde{\mathbf{H}} \succ \widetilde{\mathbf{H}}'$, 2) both are infeasible solutions and $\widetilde{\mathbf{H}}$ has a smaller constraint violation, 3) $\widetilde{\mathbf{H}}$ is feasible and $\widetilde{\mathbf{H}}'$ is infeasible, where the constraint violation ($\Delta$) of an individual $\widetilde{\mathbf{H}}$ can be calculated as follows:

$$\begin{aligned}
\Delta(\widetilde{\mathbf{H}}) = & \langle 1 - S_\Sigma | \mathbf{T} \rangle + \langle 1 - \mathfrak{C}(G_{\text{WSN}}) \rangle \\
& + \left\langle \tilde{H}_i^v - 1 \right\rangle + \left\langle \tilde{H}_i^v - \left(\tilde{H}_i^v\right)^2 - \varepsilon \right\rangle,
\end{aligned} \quad (26)$$

where $\langle \alpha \rangle$ returns a zero if $\alpha \leq 0$, otherwise it returns $|\alpha|$.

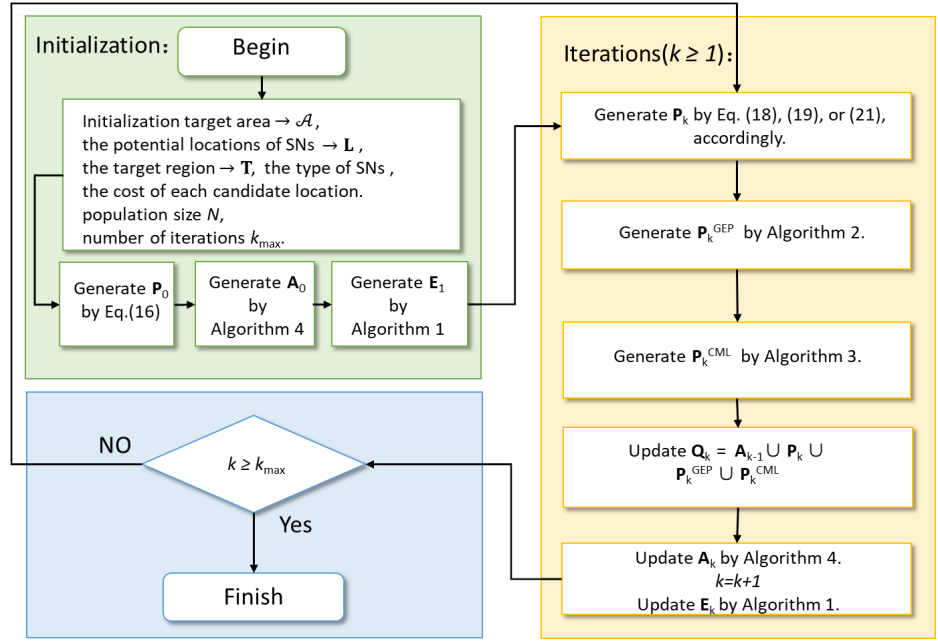

**Fig. 5** Flow chart of cost-saving WSNs deployment optimization by CMOMPA.

*4.2.3 The process of the optimal SNs deployment method*

The cost-saving optimal deployment method for WSNs based on CMOMPA is presented as follows.

**Initialization:**

Step1: Initialize the deployment area and all parameters, $k \leftarrow 0$, $N \leftarrow 100$. Randomly generate $N$ individuals (feasible solutions) to create $\mathbf{P}_k$, as modeled in Eq. (16). The $i$th individual $\widetilde{\mathbf{H}}_i$ in $\mathbf{P}_k$ is a relaxed $\mathbf{H}$ which represents a potential deployment solution. Specifically, $\widetilde{\mathbf{H}}_i$ is defined as follows:

$$\widetilde{\mathbf{H}}_i = \left[\tilde{\mathcal{H}}_1, \cdots, \tilde{\mathcal{H}}_j, \cdots, \tilde{\mathcal{H}}_{\mathbb{I}_{\text{site}}}\right]_i \quad (27)$$

where $\tilde{\mathcal{H}}_j = \left[\tilde{H}_j^1, \tilde{H}_j^2, \cdots, \tilde{H}_j^v, \cdots, \tilde{H}_j^{\mathbb{I}_{\text{type}}}\right]$.

Step2: Create $\mathbf{A}_k$ by selecting $N_a$ individuals by the reference points based method and construct $\mathbf{E}_1$ by Algorithm 1.

**Iterations**$(k \geq 1)$**:**

Step3: In the $k$th iteration, generate $\mathbf{P}_k$ by Eq. (18), (19), or (21), accordingly.

Step4: Generate $\mathbf{P}_k^{\text{GEP}}$ by Algorithm 2.

Step5: Generate $\mathbf{P}_k^{\text{CML}}$ by Algorithm 3.

Step6: Update $\mathbf{Q}_k = \mathbf{A}_{k-1} \cup \mathbf{P}_k \cup \mathbf{P}_k^{\text{GEP}} \cup \mathbf{P}_k^{\text{CML}}$.

Step7: Update $\mathbf{A}_k$ by the constrained-domination version of Algorithm 4, $k = k + 1$.

Step8: Update $\mathbf{E}_k$ by Algorithm 1.

Step9: Repeat the loop from Step 3 to 8 until the maximum number of iterations is reached.

Step10: The individuals in the final $\mathbf{A}$ are the obtain approximate Pareto optimal deployment solutions.

Fig. 5 shows the flow chart of the CMOMPA based optimal SNs deployment method.

## 5 Performance evolution

The previous section has introduced the proposed CMOMPA and the cost-saving SNs deployment method based on it. In this section, we present the performance evaluations and discuss the results. Specifically, the experimental setup, the benchmark functions, the comparison algorithms and the performance metrics are firstly introduced. Then, CMOMPA and the WSN deployment method based on it are extensively evaluated, respectively. The experimental results are discussed in detail.

5.1 Experimental setup

The performances of the proposed CMOMPA and the cost-saving SNs deployment based on it were extensively evaluated in this article. The algorithms and deployment methods were implemented in MATLAB 2020b. The performance evaluations were conducted on a server with an lntel(R) Xeon (R) E5-2620 3.0 GHz CPU, 64 GB RAM, a Windows Server 2019 operating system.

**Table 3** Detailed descriptions of benchmark functions

| Name | Features | Objectives | Dimension |
|---|---|---|---|
| ZDT1 | Unimodal, convex | 2 | 30 |
| ZDT2 | Unimodal, concave | 2 | 30 |
| ZDT3 | Multimodal, disconnected | 2 | 30 |
| ZDT4 | Multimodal, convex | 2 | 10 |
| ZDT6 | Multimodal, concave | 2 | 10 |
| DTLZ1 | Linear, multimodal | 3 | 7 |
| DTLZ2 | Concave | 3 | 12 |
| DTLZ3 | Concave, multimodal | 3 | 12 |
| DTLZ4 | Concave, biased | 3 | 12 |
| DTLZ5 | Concave, Degenerate | 3 | 12 |
| DTLZ6 | Concave, Degenerate, Biased | 3 | 12 |
| DTLZ7 | Multimodal, Mixed, Disconnected | 3 | 22 |
| WFG2 | Multimodal, nonseparable, convex | 3 | 12 |
| WFG3 | Linear, degenerate | 3 | 12 |
| WFG4 | Multimodal, concave | 3 | 12 |
| WFG5 | Concave, deceptive | 3 | 12 |
| WFG6 | Concave, nonseparable | 3 | 12 |
| WFG7 | Concave, biased | 3 | 12 |
| WFG8 | Concave, biased, nonseparable | 3 | 12 |
| WFG9 | Concave, biased, multimodal, deceptive | 3 | 12 |

5.2 Benchmark function and comparison algorithms

In this paper, 20 popular benchmark functions selected from ZDT, WFG and DTLZ benchmark suites were used to evaluate the performance of CMOMPA. The detailed descriptions of these benchmark functions are listed in Table 3. These benchmark functions have different features and can comprehensively test the performance of the algorithms. Specifically, some multimodal functions, e.g. ZDT3-6, etc., require that the algorithm has a strong global convergence ability. The functions with irregular PFs, e.g., DTLZ5-7, impose great challenges upon the diversity of algorithm. The functions in WFG suite are scaled to different ranges. These functions have high requirements for both global convergence and solution distribution of the algorithm. The maximum number of iterations for functions in the ZDT suite is 300, and that for the other suites is 3000. The population size is 100.

We compared CMOMPA with the following ten state-of-the-art multi-objective optimization algorithms, namely NSGA-II [69], non-dominated sorting genetic algorithm III (NSGA-III) [26], multi-objective evolutionary algorithm based on decomposition (MOEA/D) [70], Pareto envelope based selection II (PESA-II) [81], competitive multi-objective particle swarm optimization (CMOPSO) [82], non-dominated sorting and local search (NSLS) [83], coevolutionary for constrained multi-objective optimization (CCMO) [84], constrained multi-objective evolutionary algorithms with a two-stage (CMOEA-MS) [85], dynamically constrained NSGA-III (DCNSGA-III) [86] and MOMPA [68]. All the compared algorithms are implemented in PlatEMO [87], except for MOMPA. The detailed parameters of the algorithms are shown in Table 4. The above parameters are either exactly recommended by its developers or in the range of the recommendations to have the best performance for each algorithm.

5.3 Performance metric

In our experimental comparison, we used the inverted generational distance (IGD) [88] and the hypervolume (HV) [89] to evaluate the performance of different algorithms. The performance of the algorithm can be fully measured by the IGD and HV metrics [90, 91].

The inverted generational distance (IGD): Assuming that the approximate non-dominated solution set obtained by the multi-objective optimization algorithm is $\mathbf{P}_a$, and the reference points set sampled along the true PF is $\mathbf{P}_t$. The value of IGD can be calculated as follows.

$$\text{IGD}(\mathbf{P}_a, \mathbf{P}_t) = \frac{\sum_{\nu \in \mathbf{P}_t} \min_{\chi \in \mathbf{P}_a} \|\nu, \chi\|_2}{|\mathbf{P}_t|}, \tag{28}$$

where $|\mathbf{P}_t|$ is the size of the set $\mathbf{P}_t$ and $\|\cdot\|$ returns the Euclidean distance. Intuitively, the smaller the IGD is, the better performance is achieved. Moreover, in this paper, we use the method reported in [87] to sample reference points.

Hypervolume (HV): Let $\mathbf{z} = (z_1, z_2, \cdots, z_m)$ be a reference point, where $m$ denotes the number of objectives. Then, HV can be defined as follows.

$$\text{HV}(\mathbf{P}_a) = \text{Leb}\left(\bigcup_{\mathbf{X} \in \mathbf{P}_a} [f_1(\mathbf{X}), z_1] \times \cdots \times [f_m(\mathbf{X}), z_m]\right), \tag{29}$$

where $\text{Leb}(\cdot)$ denotes the Lebesgue measures. A higher value of HV indicates a better overall performance.

All experiments were run 30 times, independently, and the average value (Avg.) and standard deviation (Std.) of the IGD and HV were used as the final performance metric.

5.4 Performance Evaluation of CMOMPA

Table 5 and Table 6 show the comparison results on IGD and HV, respectively, among CMOMPA and other multi-objective optimization algorithms. We also rank the algorithms for each benchmark function in these two tables. It can be observed that CMOMPA is considerably promising. For IGD, CMOMPA ranks top three in 14 out of all 20 benchmark functions, namely four functions in ZDT and DTLZ suites, respectively, and six in WFG suite. For HV, it also shows strong competitiveness. In addition, Table 7 summarizes the total scores and the final rankings in terms of the integrated IGD and HV metrics. An algorithm's score is derived from its ranking on each benchmark function (as shown in Table 5 and Table 6). Concretely, the first-ranked algorithm gets one point, the second-ranked algorithm gets two points, and so on. Finally, we rank the algorithms according to the total scores. It can be found that CMOMPA also ranks the first in total score with 169 points.

**Table 4** The parameters of comparative algorithms.

| Algorithm | Parameters | Value |
| --- | --- | --- |
| NSGA-II, NSGA-III | Crossover probability | 1 |
| DCNSGA-III, CCMO | Mutation probability | 1/D [1] |
| CMOEA-MS | Distribution indexes for SBX | 20 |
| PESA-II, MOEA/D | Distribution indexes for polynomial mutation | 20 |
| MOEA/D | Penalty parameter of the PBI function | 5 |
|  | Neighbourhood size | Pop-size/10 [2] |
| CMOPSO | Size of elite particle set | 10 |
| NSLS | Local Search rates | $C_n \sim N(0.5, 0.1)$ |
| MOMPA | $\theta$ | 0.5 |
|  | $F_A$ | 0.2 |
| CMOMPA | $\theta$ | 0.5 |
|  | Distribution indexes for polynomial mutation | 20 |
| DCNSGA-III | $cp$ | 5 |
| CMOEA-MS | $\lambda$ | 0.5 |

[1] D denotes the dimensionality of the problem decision variable.
[2] Pop-size denotes population size.

The reasons for these results can be interpreted twofold. Firstly, benefiting from the competition mechanism-based learning strategy, CMOMPA has a considerably high global and local search performance. This makes CMOMPA outperforms MOMPA, which ranks the five in total score and adopts the same elite selecting method as CMOMPA. Secondly, the results indicate that for DTLZ1 and DTLZ3, CMOMPA can jump out of the optimal local solution and approach the true PF. However, for benchmark functions with irregular PFs, such as DTLZ5-7, the performance of CMOMPA algorithm degrades. This may be because the solutions selection method based on uniformly distributed reference points is not very suitable for such problems. CMOMPA also shows unsatisfactory performance in WFG3 and WFG6. Since the PF of WFG3 is a degenerated curve, competition mechanism based on learning strategy may generate too many uncertain solutions which result in poor performance. WFG6 is a nonseparable-reduced problem. NSLS has strong local exploration ability and can deal with WFG6 well. Compared with NSLS, CMOMPA's local exploration capability needs to be further strengthened to more effectively handle such problems.

The Wilcoxon signed-rank test method [92] was taken to evaluate the significance of the involved algorithms. From the Wilcoxon signed-rank test results of the IGD shown in Table 8, we can see that most of the p-values obtained by comparing CMOMPA with other multi-objective optimization algorithms are much less than 0.05, which demonstrates significant differences between them in a statistical sense. In addition, in the last row, the number of plus signs, minus signs and equal signs represent the number of benchmark functions in which CMOMPA is superior to, inferior to and non-significantly different from the counterpart multi-objective optimization algorithms, respectively. It can be seen that there are always more plus signs than minus signs, which indicates that CMOMPA has a significant statistical superiority compared to other algorithms.

To investigate the sensitivity of the tuning probability $\theta$ which can attenuate/magnify the individual optimization step in CMOMPA, we separately tested the convergence performance of CMOMPA with different values of $\theta$, i.e., $\theta = \{0.1, 0.3, 0.5, 0.7, 0.9\}$ on ZDT6, DTLZ1, WFG7. Other parameter settings are consistent with the previous ones. The results were averaged from 30 independent runs. As can be seen from Fig. 6, for three different functions, the IGD value of CMOMPA finally converges to the same level, which indicates that CMOMPA is not sensitive to the parameter $\theta$.

**Table 5** Comparisions on computational results among CMOMPA and other multi-objective optimization algorithms about the IGD metric.

| Algrithm | ZDT1 Avg. | Std. | Rank | ZDT2 Avg. | Std. | Rank | ZDT3 Avg. | Std. | Rank | ZDT4 Avg. | Std. | Rank |
|---|---|---|---|---|---|---|---|---|---|---|---|---|
| NSGA-II | 4.82E-03 | 1.75E-04 | 7 | 4.86E-03 | 1.94E-04 | 8 | 7.35E-03 | 7.40E-03 | 7 | 5.41E-03 | 8.81E-04 | 3 |
| NSGA-III | 3.91E-03 | 1.20E-05 | 4 | 3.86E-03 | 2.78E-05 | 4 | 7.03E-03 | 5.23E-03 | 6 | 1.28E-02 | 1.73E-02 | 7 |
| MOEA/D | 1.19E-02 | 8.49E-03 | 10 | 2.60E-02 | 2.70E-02 | 10 | 3.03E-02 | 2.10E-02 | 10 | 2.09E-02 | 1.24E-02 | 9 |
| PESA-II | 1.15E-02 | 3.44E-03 | 9 | 1.14E-02 | 1.83E-03 | 9 | 2.06E-02 | 1.42E-02 | 9 | 1.34E-02 | 3.10E-03 | 8 |
| CMOPSO | 4.19E-03 | 8.96E-05 | 6 | 4.13E-03 | 8.86E-05 | 7 | **4.64E-03** | 6.08E-05 | 1 | 2.60E-01 | 2.54E-01 | 10 |
| NSLS | 2.34E-01 | 2.42E-02 | 11 | 4.05E-01 | 5.69E-02 | 11 | 2.27E-01 | 3.81E-02 | 11 | 8.26E-01 | 2.35E-01 | 11 |
| CCMO | 3.96E-03 | 7.71E-05 | 5 | 3.95E-03 | 6.06E-05 | 5 | 4.94E-03 | 1.22E-04 | 2 | 4.98E-03 | 1.11E-03 | 2 |
| CMOEA-MS | 4.85E-03 | 1.78E-04 | 8 | 3.97E-03 | 5.90E-05 | 6 | 9.74E-03 | 9.84E-04 | 8 | 6.14E-03 | 1.57E-03 | 4 |
| DCNSGA-III | 3.90E-03 | 1.08E-05 | 3 | 3.83E-03 | 1.68E-05 | 3 | 6.42E-03 | 5.33E-03 | 5 | 6.61E-03 | 4.50E-03 | 5 |
| MOMPA | 3.90E-03 | 7.58E-05 | 2 | 3.80E-03 | 8.92E-06 | 1 | 6.30E-03 | 3.16E-04 | 4 | **3.90E-03** | 5.87E-05 | 1 |
| CMOMPA | **3.90E-03** | 3.03E-05 | 1 [1] | **3.80E-03** | 8.92E-06 | 1 | 6.11E-03 | 2.63E-04 | 3 | 7.64E-03 | 1.80E-02 | 6 |

| Algrithm | ZDT6 Avg. | Std. | Rank | WFG2 Avg. | Std. | Rank | WFG3 Avg. | Std. | Rank | WFG4 Avg. | Std. | Rank |
|---|---|---|---|---|---|---|---|---|---|---|---|---|
| NSGA-II | 3.72E-03 | 1.13E-04 | 8 | 2.19E-01 | 9.86E-03 | 10 | 8.80E-02 | 1.40E-02 | 2 | 2.73E-01 | 9.96E-03 | 10 |
| NSGA-III | 3.21E-03 | 2.75E-04 | 6 | 1.66E-01 | 1.06E-03 | 4 | 9.81E-02 | 1.10E-02 | 5 | 2.21E-01 | 3.05E-05 | 6 |
| MOEA/D | 7.00E-03 | 1.16E-03 | 10 | 2.19E-01 | 2.23E-02 | 11 | 1.57E-01 | 3.79E-04 | 8 | 2.47E-01 | 2.09E-03 | 8 |
| PESA-II | 7.43E-03 | 7.99E-04 | 11 | 1.93E-01 | 8.62E-03 | 8 | 4.09E-01 | 1.90E-01 | 11 | 2.93E-01 | 1.57E-02 | 11 |
| CMOPSO | 3.11E-03 | 2.76E-05 | 3 | 1.82E-01 | 6.33E-03 | 7 | 1.47E-01 | 1.30E-02 | 7 | 2.60E-01 | 4.05E-03 | 9 |
| NSLS | 6.23E-03 | 1.87E-03 | 9 | 1.74E-01 | 4.18E-03 | 5 | **8.25E-02** | 1.69E-02 | 1 | **2.13E-01** | 2.40E-03 | 1 |
| CCMO | 3.15E-03 | 7.51E-05 | 4 | 1.75E-01 | 5.23E-03 | 6 | 9.76E-02 | 7.54E-03 | 4 | 2.16E-01 | 2.76E-03 | 4 |
| CMOEA-MS | 3.17E-03 | 5.92E-05 | 5 | 2.03E-01 | 1.35E-02 | 9 | 1.72E-01 | 1.94E-02 | 10 | 2.45E-01 | 5.73E-03 | 7 |
| DCNSGA-III | 3.37E-03 | 4.23E-04 | 7 | 1.65E-01 | 1.27E-03 | 3 | 9.86E-02 | 1.11E-02 | 6 | 2.21E-01 | 1.60E-05 | 5 |
| MOMPA | 3.00E-03 | 1.01E-05 | 2 | **1.57E-01** | 2.34E-03 | 1 | 9.11E-02 | 1.08E-02 | 3 | 2.14E-01 | 8.59E-04 | 3 |
| CMOMPA | **2.99E-03** | 8.01E-06 | 1 | 1.58E-01 | 2.28E-03 | 2 | 1.66E-01 | 2.39E-02 | 9 | 2.14E-01 | 5.92E-04 | 2 |

| Algrithm | WFG5 Avg. | Std. | Rank | WFG6 Avg. | Std. | Rank | WFG7 Avg. | Std. | Rank | WFG8 Avg. | Std. | Rank |
|---|---|---|---|---|---|---|---|---|---|---|---|---|
| NSGA-II | 2.80E-01 | 9.50E-03 | 11 | 3.03E-01 | 1.80E-02 | 10 | 2.83E-01 | 1.16E-02 | 10 | 3.74E-01 | 1.04E-02 | 10 |
| NSGA-III | 2.30E-01 | 9.22E-06 | 5 | 2.34E-01 | 6.99E-03 | 2 | 2.21E-01 | 1.67E-05 | 4 | 2.78E-01 | 3.42E-03 | 3 |
| MOEA/D | 2.47E-01 | 1.76E-03 | 7 | 2.68E-01 | 1.11E-02 | 7 | 2.44E-01 | 1.61E-03 | 7 | 2.97E-01 | 2.00E-03 | 7 |
| PESA-II | 2.78E-01 | 9.47E-03 | 10 | 3.11E-01 | 1.58E-02 | 11 | 2.87E-01 | 1.36E-02 | 11 | 3.78E-01 | 1.61E-02 | 11 |
| CMOPSO | 2.50E-01 | 5.05E-03 | 8 | 2.37E-01 | 4.58E-03 | 3 | 2.33E-01 | 4.65E-03 | 6 | 3.31E-01 | 5.59E-03 | 9 |
| NSLS | **2.16E-01** | 2.24E-03 | 1 | **2.15E-01** | 3.04E-03 | 1 | 2.70E-01 | 7.57E-03 | 9 | 2.84E-01 | 4.49E-03 | 4 |
| CCMO | 2.22E-01 | 2.11E-03 | 4 | 2.39E-01 | 9.46E-03 | 5 | 2.17E-01 | 2.64E-03 | 3 | 2.96E-01 | 4.01E-03 | 6 |
| CMOEA-MS | 2.53E-01 | 5.05E-03 | 9 | 2.65E-01 | 9.89E-03 | 6 | 2.48E-01 | 5.68E-03 | 8 | 3.25E-01 | 6.35E-03 | 8 |
| DCNSGA-III | 2.30E-01 | 5.86E-06 | 6 | 2.38E-01 | 7.04E-03 | 4 | 2.21E-01 | 1.30E-05 | 5 | **2.75E-01** | 3.25E-03 | 1 |
| MOMPA | 2.21E-01 | 3.50E-03 | 2 | 2.70E-01 | 6.02E-02 | 8 | 2.14E-01 | 7.35E-04 | 2 | 2.87E-01 | 4.10E-03 | 5 |
| CMOMPA | 2.22E-01 | 1.24E-03 | 3 | 2.77E-01 | 6.41E-02 | 9 | **2.14E-01** | 6.65E-04 | 1 | 2.77E-01 | 1.00E-02 | 2 |

| Algrithm | WFG9 Avg. | Std. | Rank | DTLZ1 Avg. | Std. | Rank | DTLZ2 Avg. | Std. | Rank | DTLZ3 Avg. | Std. | Rank |
|---|---|---|---|---|---|---|---|---|---|---|---|---|
| NSGA-II | 2.76E-01 | 1.39E-02 | 11 | 2.74E-02 | 1.34E-03 | 10 | 6.94E-02 | 2.62E-03 | 11 | 6.83E-02 | 2.90E-03 | 7 |
| NSGA-III | 2.21E-01 | 5.77E-04 | 5 | 2.06E-02 | 2.71E-06 | 4 | 5.45E-02 | 7.63E-07 | 6 | 5.45E-02 | 8.71E-06 | 4 |
| MOEA/D | 2.48E-01 | 2.03E-02 | 9 | 2.06E-02 | 6.79E-07 | 3 | 5.45E-02 | 5.07E-08 | 6 | 5.45E-02 | 1.50E-05 | 6 |
| PESA-II | 2.76E-01 | 2.69E-02 | 10 | 2.47E-02 | 1.43E-03 | 9 | 6.73E-02 | 3.70E-03 | 10 | 7.18E-02 | 9.97E-03 | 8 |
| CMOPSO | 2.19E-01 | 3.35E-03 | 4 | 2.07E-02 | 3.74E-04 | 6 | 5.76E-02 | 9.24E-04 | 9 | 3.66E+00 | 3.92E+00 | 11 |
| NSLS | 2.40E-01 | 5.68E-03 | 7 | 2.39E-01 | 1.69E-01 | 11 | 5.42E-02 | 7.03E-04 | 4 | 2.63E+00 | 1.21E+00 | 9 |
| CCMO | 2.16E-01 | 4.08E-03 | 3 | 2.01E-02 | 1.28E-04 | 2 | 5.43E-02 | 4.71E-04 | 5 | 5.34E-02 | 6.14E-04 | 2 |
| CMOEA-MS | 2.46E-01 | 2.18E-02 | 8 | 2.09E-02 | 3.04E-04 | 7 | 5.39E-02 | 4.90E-04 | 3 | **5.32E-02** | 6.52E-04 | 1 |
| DCNSGA-III | 2.25E-01 | 2.19E-02 | 6 | 2.06E-02 | 2.85E-06 | 4 | 5.45E-02 | 1.28E-06 | 6 | 5.45E-02 | 2.15E-05 | 5 |
| MOMPA | **2.12E-01** | 5.32E-04 | 1 | 2.09E-02 | 8.83E-04 | 8 | **5.25E-02** | 2.32E-05 | 1 | 2.80E+00 | 9.99E-01 | 10 |
| CMOMPA | 2.13E-01 | 6.27E-04 | 2 | **2.01E-02** | 3.38E-05 | 1 | 5.35E-02 | 9.89E-04 | 2 | 5.36E-02 | 3.74E-04 | 3 |

| Algrithm | DTLZ4 Avg. | Std. | Rank | DTLZ5 Avg. | Std. | Rank | DTLZ6 Avg. | Std. | Rank | DTLZ7 Avg. | Std. | Rank |
|---|---|---|---|---|---|---|---|---|---|---|---|---|
| NSGA-II | 6.82E-02 | 2.58E-03 | 4 | 5.85E-03 | 3.16E-04 | 5 | 5.82E-03 | 2.88E-04 | 5 | 7.77E-02 | 5.40E-03 | 6 |
| NSGA-III | 1.20E-01 | 1.68E-01 | 7 | 1.22E-02 | 1.57E-03 | 8 | 1.83E-02 | 2.36E-03 | 9 | 7.73E-02 | 3.83E-03 | 5 |
| MOEA/D | 5.45E-02 | 7.92E-06 | 3 | 3.39E-02 | 2.11E-06 | 11 | 3.39E-02 | 5.22E-07 | 11 | 1.75E-01 | 1.18E-01 | 11 |
| PESA-II | 9.44E-02 | 1.61E-01 | 6 | 1.22E-02 | 3.03E-03 | 7 | 1.39E-02 | 2.75E-03 | 6 | 1.43E-01 | 1.66E-01 | 10 |
| CMOPSO | 2.09E-01 | 3.35E-01 | 11 | 5.21E-03 | 4.95E-04 | 4 | 4.20E-03 | 4.83E-05 | 3 | 8.49E-02 | 7.12E-02 | 7 |
| NSLS | 1.54E-01 | 1.09E-01 | 9 | 4.50E-03 | 6.88E-05 | 3 | 4.48E-03 | 4.62E-05 | 4 | **6.04E-02** | 1.34E-03 | 1 |
| CCMO | 1.65E-01 | 2.36E-01 | 10 | 4.34E-03 | 8.17E-05 | 2 | **4.09E-03** | 2.75E-05 | 1 | 6.06E-02 | 1.24E-03 | 2 |
| CMOEA-MS | 8.66E-02 | 1.23E-01 | 5 | **4.31E-03** | 9.15E-05 | 1 | 4.09E-03 | 2.85E-05 | 2 | 7.59E-02 | 5.10E-03 | 3 |
| DCNSGA-III | 1.36E-01 | 1.85E-01 | 8 | 1.19E-02 | 1.69E-03 | 6 | 1.79E-02 | 2.62E-03 | 8 | 7.65E-02 | 2.64E-03 | 4 |
| MOMPA | **5.25E-02** | 3.64E-05 | 1 | 1.78E-02 | 2.27E-03 | 10 | 2.32E-02 | 2.85E-03 | 10 | 9.09E-02 | 8.21E-03 | 9 |
| CMOMPA | 5.29E-02 | 5.01E-04 | 2 | 1.47E-02 | 1.58E-03 | 9 | 1.70E-02 | 2.50E-03 | 7 | 8.55E-02 | 4.70E-03 | 8 |

[1] The Avg. of CMOMPA, MOMPA, DCNSGA-III are specified as 0.003900, 0.003900, 0.003903

5.5 Performance evaluation of the CMOMPA based deployment method

Suppose a $55 \times 55 \times 20$ m$^3$ three-storey factory area, as shown in Fig. 7, the location set **L** where nodes can be deployed is known. The deployment cost of each candidate location ($C_{\text{loc}}(i), i = \{1, 2, \cdots, \mathbb{I}_{\text{site}}\}$) is assigned with a random integer in $[1, 5]$, and only one

**Table 6** Comparisions on computational results among CMOMPA and other multi-objective optimization algorithms about the HV metric.

| Algrithm | ZDT1 | | | ZDT2 | | | ZDT3 | | | ZDT4 | | |
|---|---|---|---|---|---|---|---|---|---|---|---|---|
| | Avg. | Std. | Rank | Avg. | Std. | Rank | Avg. | Std. | Rank | Avg. | Std. | Rank |
| NSGA-II | 0.719120 | 0.000266 | 7 | 0.443780 | 0.000266 | 8 | 0.605250 | 0.022500 | 3 | 0.717000 | 0.001860 | 5 |
| NSGA-III | 0.720090 | 0.000070 | 4 | 0.444780 | 0.000114 | 3 | 0.601640 | 0.016200 | 4 | 0.717170 | 0.002330 | 4 |
| MOEA/D | 0.713050 | 0.004180 | 9 | 0.404350 | 0.034600 | 10 | **0.621210** | 0.044700 | 1 | 0.692720 | 0.014700 | 9 |
| PESA-II | 0.711580 | 0.002160 | 10 | 0.432750 | 0.004460 | 9 | 0.618980 | 0.038200 | 2 | 0.707810 | 0.002970 | 8 |
| CMOPSO | 0.719300 | 0.000267 | 6 | 0.444000 | 0.000250 | 7 | 0.599680 | 0.000114 | 5 | 0.379460 | 0.209000 | 10 |
| NSLS | 0.406270 | 0.021600 | 11 | 0.055655 | 0.026500 | 11 | 0.486210 | 0.047100 | 11 | 0.026129 | 0.054900 | 11 |
| CCMO | 0.720080 | 0.000109 | 5 | 0.444700 | 0.000099 | 5 | 0.599510 | 0.000085 | 6 | 0.717320 | 0.001710 | 3 |
| CMOEA-MS | 0.718550 | 0.000352 | 8 | 0.444670 | 0.000143 | 6 | 0.598470 | 0.016900 | 10 | 0.714590 | 0.002630 | 7 |
| DCNSGA-III | 0.720100 | 0.000069 | 2 | 0.444750 | 0.000098 | 4 | 0.599370 | 0.000079 | 7 | 0.715740 | 0.004310 | 6 |
| MOMPA | **0.720100** | 0.000049 | 1 | 0.444800 | 0.000050 | 2 | 0.598600 | 0.000319 | 9 | **0.719900** | 0.000181 | 1 |
| CMOMPA | 0.720100 | 0.000095 | 3 | **0.444900** | 0.000031 | 1 | 0.598800 | 0.000324 | 8 | 0.718500 | 0.003800 | 2 |

| Algrithm | ZDT6 | | | WFG2 | | | WFG3 | | | WFG4 | | |
|---|---|---|---|---|---|---|---|---|---|---|---|---|
| | Avg. | Std. | Rank | Avg. | Std. | Rank | Avg. | Std. | Rank | Avg. | Std. | Rank |
| NSGA-II | 0.388000 | 0.000255 | 8 | 0.920290 | 0.002920 | 10 | **0.402030** | 0.002810 | 1 | 0.523790 | 0.003830 | 9 |
| NSGA-III | 0.388300 | 0.000330 | 6 | **0.931080** | 0.000694 | 1 | 0.391970 | 0.004610 | 4 | **0.559620** | 0.000005 | 1 |
| MOEA/D | 0.381030 | 0.002080 | 11 | 0.921670 | 0.004390 | 9 | 0.365400 | 0.000264 | 7 | 0.549110 | 0.000763 | 5 |
| PESA-II | 0.383790 | 0.000883 | 10 | 0.922080 | 0.002210 | 8 | 0.260260 | 0.049400 | 11 | 0.476690 | 0.007260 | 11 |
| CMOPSO | 0.388850 | 0.000039 | 3 | 0.929210 | 0.001550 | 5 | 0.359250 | 0.005670 | 9 | 0.493550 | 0.006100 | 10 |
| NSLS | 0.386510 | 0.001200 | 9 | 0.922570 | 0.008260 | 7 | 0.390330 | 0.007800 | 5 | 0.557100 | 0.001730 | 3 |
| CCMO | 0.388330 | 0.000406 | 5 | 0.930600 | 0.000938 | 3 | 0.389370 | 0.003700 | 6 | 0.543480 | 0.002680 | 7 |
| CMOEA-MS | 0.388400 | 0.000323 | 4 | 0.914740 | 0.004380 | 11 | 0.346940 | 0.011300 | 10 | 0.535970 | 0.003030 | 8 |
| DCNSGA-III | 0.388300 | 0.000534 | 7 | 0.930420 | 0.000628 | 2 | 0.393480 | 0.004220 | 3 | 0.559610 | 0.000011 | 2 |
| MOMPA | **0.389000** | 0.000001 | 1 | 0.928300 | 0.001100 | 6 | 0.395700 | 0.003800 | 2 | 0.548000 | 0.001100 | 6 |
| CMOMPA | 0.389000 | 0.000003 | 2 | 0.929800 | 0.001000 | 4 | 0.361300 | 0.009800 | 8 | 0.552400 | 0.000815 | 4 |

| Algrithm | WFG5 | | | WFG6 | | | WFG7 | | | WFG8 | | |
|---|---|---|---|---|---|---|---|---|---|---|---|---|
| | Avg. | Std. | Rank | Avg. | Std. | Rank | Avg. | Std. | Rank | Avg. | Std. | Rank |
| NSGA-II | 0.493290 | 0.004160 | 9 | 0.472760 | 0.013200 | 10 | 0.521400 | 0.003970 | 9 | 0.444410 | 0.002790 | 9 |
| NSGA-III | 0.518470 | 0.000060 | 2 | 0.508340 | 0.012600 | 3 | 0.559380 | 0.000060 | 2 | **0.477600** | 0.001830 | 1 |
| MOEA/D | 0.506420 | 0.004740 | 7 | 0.504830 | 0.014800 | 5 | 0.540810 | 0.002130 | 6 | 0.471440 | 0.001940 | 6 |
| PESA-II | 0.470630 | 0.006490 | 11 | 0.456990 | 0.015900 | 11 | 0.483430 | 0.010500 | 11 | 0.408740 | 0.006860 | 11 |
| CMOPSO | 0.478590 | 0.004850 | 10 | 0.522890 | 0.003280 | 2 | 0.525330 | 0.003220 | 8 | 0.439430 | 0.004490 | 10 |
| NSLS | 0.507120 | 0.001160 | 6 | **0.545720** | 0.005990 | 1 | 0.498660 | 0.005100 | 10 | 0.471630 | 0.002180 | 5 |
| CCMO | 0.513740 | 0.001590 | 5 | 0.498820 | 0.011500 | 6 | 0.543630 | 0.001990 | 5 | 0.458790 | 0.002090 | 7 |
| CMOEA-MS | 0.504470 | 0.002170 | 8 | 0.496590 | 0.013200 | 8 | 0.535470 | 0.003260 | 7 | 0.444680 | 0.003500 | 8 |
| DCNSGA-III | **0.518470** | 0.000007 | 1 | 0.507160 | 0.010900 | 4 | **0.559380** | 0.000048 | 1 | 0.476950 | 0.001730 | 2 |
| MOMPA | 0.518100 | 0.000109 | 4 | 0.478300 | 0.069200 | 9 | 0.552400 | 0.000617 | 4 | 0.476100 | 0.003000 | 3 |
| CMOMPA | 0.518300 | 0.000262 | 3 | 0.498400 | 0.069500 | 7 | 0.555100 | 0.000530 | 3 | 0.475400 | 0.005800 | 4 |

| Algrithm | WFG9 | | | DTLZ1 | | | DTLZ2 | | | DTLZ3 | | |
|---|---|---|---|---|---|---|---|---|---|---|---|---|
| | Avg. | Std. | Rank | Avg. | Std. | Rank | Avg. | Std. | Rank | Avg. | Std. | Rank |
| NSGA-II | 0.505850 | 0.006110 | 9 | 0.824360 | 0.003310 | 10 | 0.533300 | 0.003630 | 10 | 0.535290 | 0.004820 | 7 |
| NSGA-III | 0.545950 | 0.001400 | 3 | 0.841720 | 0.000025 | 3 | 0.559610 | 0.000014 | 4 | 0.559350 | 0.000273 | 3 |
| MOEA/D | 0.514900 | 0.024100 | 7 | 0.841720 | 0.000026 | 4 | 0.559620 | 0.000000 | 2 | 0.559210 | 0.000818 | 4 |
| PESA-II | 0.482460 | 0.005090 | 11 | 0.824470 | 0.005570 | 9 | 0.518510 | 0.008710 | 11 | 0.526740 | 0.010800 | 8 |
| CMOPSO | 0.519540 | 0.003450 | 5 | 0.839710 | 0.001270 | 6 | 0.541570 | 0.003010 | 9 | 0.210420 | 0.258000 | 9 |
| NSLS | 0.494660 | 0.003460 | 10 | 0.358150 | 0.328000 | 11 | **0.561700** | 0.000601 | 1 | 0.018735 | 0.103000 | 11 |
| CCMO | 0.517880 | 0.004680 | 6 | **0.842580** | 0.000289 | 1 | 0.555090 | 0.001130 | 7 | **0.560850** | 0.000908 | 1 |
| CMOEA-MS | 0.510630 | 0.007010 | 8 | 0.839610 | 0.001180 | 7 | 0.556470 | 0.000996 | 6 | 0.560460 | 0.001330 | 2 |
| DCNSGA-III | 0.544940 | 0.002440 | 4 | 0.841730 | 0.000008 | 2 | 0.559620 | 0.000005 | 3 | 0.559160 | 0.000552 | 5 |
| MOMPA | 0.549500 | 0.001300 | 2 | 0.832700 | 0.020000 | 8 | 0.558100 | 0.000209 | 5 | 0.030400 | 0.117100 | 10 |
| CMOMPA | **0.549500** | 0.000839 | 1 | 0.841000 | 0.000258 | 5 | 0.554000 | 0.004400 | 8 | 0.553800 | 0.011600 | 6 |

| Algrithm | DTLZ4 | | | DTLZ5 | | | DTLZ6 | | | DTLZ7 | | |
|---|---|---|---|---|---|---|---|---|---|---|---|---|
| | Avg. | Std. | Rank | Avg. | Std. | Rank | Avg. | Std. | Rank | Avg. | Std. | Rank |
| NSGA-II | 0.505170 | 0.113000 | 10 | 0.199120 | 0.000226 | 4 | 0.199460 | 0.000115 | 5 | 0.268180 | 0.001970 | 7 |
| NSGA-III | 0.537400 | 0.067300 | 5 | 0.193940 | 0.001270 | 6 | 0.191140 | 0.001560 | 8 | 0.269340 | 0.002040 | 5 |
| MOEA/D | 0.545070 | 0.055400 | 3 | 0.181850 | 0.000001 | 11 | 0.181850 | 0.000000 | 11 | 0.257790 | 0.000301 | 11 |
| PESA-II | 0.537270 | 0.004510 | 6 | 0.192430 | 0.003400 | 8 | 0.190560 | 0.005580 | 9 | 0.261110 | 0.013200 | 10 |
| CMOPSO | 0.488950 | 0.135000 | 11 | 0.198710 | 0.000373 | 5 | **0.200180** | 0.000040 | 1 | 0.263970 | 0.015300 | 8 |
| NSLS | 0.541370 | 0.038400 | 4 | **0.199850** | 0.000073 | 1 | 0.199890 | 0.000088 | 4 | **0.278330** | 0.000823 | 1 |
| CCMO | 0.519960 | 0.078600 | 9 | 0.199680 | 0.000106 | 3 | 0.200060 | 0.000038 | 2 | 0.277130 | 0.000947 | 2 |
| CMOEA-MS | 0.534630 | 0.063500 | 7 | 0.199700 | 0.000112 | 2 | 0.200060 | 0.000039 | 3 | 0.275140 | 0.001130 | 3 |
| DCNSGA-III | 0.529360 | 0.099600 | 8 | 0.193860 | 0.001270 | 7 | 0.191290 | 0.001420 | 7 | 0.271500 | 0.005970 | 4 |
| MOMPA | **0.557400** | 0.000462 | 1 | 0.191300 | 0.001800 | 10 | 0.190100 | 0.001800 | 10 | 0.268400 | 0.003200 | 6 |
| CMOMPA | 0.555500 | 0.001100 | 2 | 0.191300 | 0.001300 | 9 | 0.192900 | 0.001300 | 6 | 0.262100 | 0.003400 | 9 |

**Table 7** Scores and ranking of various algorithms about the IGD and HV.

| Algorithms | CMOMPA | CCMO | NSGA-III | DCNSGA-III | MOMPA | CMOEA-MS | NSLS | CMOPSO | MOEA/D | NSGA-II | PESA-II |
|---|---|---|---|---|---|---|---|---|---|---|---|
| Scores of IGD | 74 | 77 | 104 | 100 | 84 | 118 | 123 | 131 | 164 | 155 | 185 |
| Scores of HV | 95 | 94 | 72 | 81 | 100 | 133 | 133 | 139 | 138 | 150 | 185 |
| Total | 169 | 171 | 176 | 181 | 184 | 251 | 256 | 270 | 302 | 305 | 370 |
| Ranking | 1 | 2 | 3 | 4 | 5 | 6 | 7 | 8 | 9 | 10 | 11 |

**Table 8** P-values of Wilcoxon test between CMOMPA and other multi-objective optimization algorithms about the IGD metric.

| | NSGA-II | NSGA-III | MOEA/D | PESA-II | CMOPSO | NSLS | MOMPA | CCMO | CMOEA-MS | DCNSGA-III |
|---|---|---|---|---|---|---|---|---|---|---|
| ZDT1 | 3.02E-11(+) | 1.10E-08(+) | 3.02E-11(+) | 3.02E-11(+) | 2.23E-09(+) | 3.02E-11(+) | 3.02E-1(+) | 7.22E-06(+) | 3.02E-11(+) | 4.62E-10(+) |
| ZDT2 | 3.02E-11(+) | 3.02E-11(+) | 3.02E-11(+) | 3.02E-11(+) | 3.02E-11(+) | 3.02E-11(+) | 3.02E-11(+) | 3.02E-11(+) | 3.02E-11(+) | 9.92E-11(+) |
| ZDT3 | 5.09E-08(+) | 6.84E-01(=) | 3.02E-11(+) | 3.02E-11(+) | 3.02E-11(-) | 3.02E-11(+) | 3.02E-11(+) | 3.02E-11(-) | 3.02E-11(+) | 3.50E-09(+) |
| ZDT4 | 6.05E-07(-) | 1.73E-07(+) | 7.38E-10(+) | 2.23E-09(+) | 2.37E-10(+) | 3.02E-11(+) | 5.57E-10(-) | 2.88E-06(-) | 3.01E-07(-) | 3.26E-07(-) |
| ZDT6 | 3.02E-11(+) | 3.69E-11(+) | 3.02E-11(+) | 3.02E-11(+) | 3.02E-11(+) | 3.02E-11(+) | 3.02E-11(+) | 3.02E-11(+) | 3.02E-11(+) | 3.02E-11(+) |
| WFG2 | 3.02E-11(+) | 7.39E-11(+) | 3.02E-11(+) | 3.02E-11(+) | 3.02E-11(+) | 3.34E-11(+) | 4.40E-03(-) | 3.02E-11(+) | 3.02E-11(+) | 1.21E-10(+) |
| WFG3 | 3.02E-11(-) | 3.69E-11(-) | 3.99E-04(-) | 1.20E-08(+) | 4.64E-05(-) | 3.02E-11(-) | 3.02E-11(-) | 3.02E-11(-) | 5.11E-01(=) | 4.50E-11(-) |
| WFG4 | 3.02E-11(+) | 3.02E-11(+) | 3.02E-11(+) | 3.02E-11(+) | 3.02E-11(+) | 5.30E-11(+) | 3.02E-11(+) | 4.80E-07(+) | 3.02E-11(+) | 3.02E-11(+) |
| WFG5 | 3.02E-11(+) | 3.02E-11(+) | 3.02E-11(+) | 3.02E-11(+) | 3.02E-11(+) | 2.15E-10(-) | 3.02E-11(-) | 4.92E-01(=) | 3.02E-11(+) | 3.02E-11(+) |
| WFG6 | 5.90E-01(=) | 7.85E-01(=) | 6.63E-01(=) | 5.20E-01(=) | 6.74E-01(=) | 1.76E-02(-) | 3.02E-11(-) | 7.17E-01(=) | 6.63E-01(=) | 7.39E-01(=) |
| WFG7 | 3.02E-11(+) | 3.02E-11(+) | 3.02E-11(+) | 3.02E-11(+) | 3.02E-11(+) | 3.02E-11(+) | 3.02E-11(+) | 1.29E-06(+) | 3.02E-11(+) | 3.02E-11(+) |
| WFG8 | 3.02E-11(+) | 7.85E-01(=) | 5.07E-10(+) | 3.02E-11(+) | 3.02E-11(+) | 2.10E-03(+) | 3.02E-11(+) | 6.12E-10(+) | 3.02E-11(+) | 1.45E-01(=) |
| WFG9 | 3.02E-11(+) | 3.02E-11(+) | 3.02E-11(+) | 3.02E-11(+) | 1.17E-09(+) | 3.02E-11(+) | 3.02E-11(-) | 1.63E-02(+) | 3.02E-11(+) | 3.02E-11(+) |
| DTLZ1 | 3.02E-11(+) | 3.02E-11(+) | 3.02E-11(+) | 3.02E-11(+) | 4.08E-11(+) | 4.50E-11(+) | 3.02E-11(+) | 6.20E-01(=) | 3.02E-11(+) | 3.02E-11(+) |
| DTLZ2 | 3.02E-11(+) | 6.77E-05(+) | 6.77E-05(+) | 3.02E-11(+) | 4.08E-11(+) | 1.40E-03(+) | 3.02E-11(-) | 2.84E-04(+) | 6.10E-03(+) | 6.77E-05(+) |
| DTLZ3 | 3.02E-11(+) | 5.57E-10(+) | 5.57E-10(+) | 3.02E-11(+) | 3.02E-11(+) | 3.02E-11(+) | 3.02E-11(-) | 1.81E-01(=) | 4.90E-03(-) | 5.57E-10(+) |
| DTLZ4 | 3.02E-11(+) | 2.44E-09(+) | 8.48E-09(+) | 3.02E-11(+) | 2.98E-11(+) | 2.37E-10(+) | 3.02E-11(-) | 6.72E-10(+) | 3.50E-09(+) | 3.50E-09(+) |
| DTLZ5 | 3.02E-11(-) | 1.11E-06(-) | 3.02E-11(+) | 8.88E-06(+) | 3.02E-11(-) | 3.02E-11(-) | 3.02E-11(+) | 3.02E-11(-) | 3.02E-11(-) | 1.60E-07(-) |
| DTLZ6 | 3.02E-11(+) | 2.15E-02(+) | 3.02E-11(+) | 8.88E-06(+) | 3.02E-11(+) | 3.02E-11(-) | 3.02E-11(+) | 3.02E-11(-) | 3.02E-11(-) | 1.26E-01(=) |
| DTLZ7 | 4.80E-07(-) | 2.39E-08(-) | 3.02E-11(+) | 6.77E-05(+) | 8.48E-09(-) | 3.02E-11(-) | 3.02E-11(+) | 3.02E-11(-) | 5.57E-10(-) | 4.62E-10(-) |
| +\=\- | 14\1\5 | 14\3\3 | 18\1\1 | 17\1\2 | 14\1\5 | 13\1\6 | 12\0\8 | 10\4\6 | 13\2\5 | 13\3\4 |

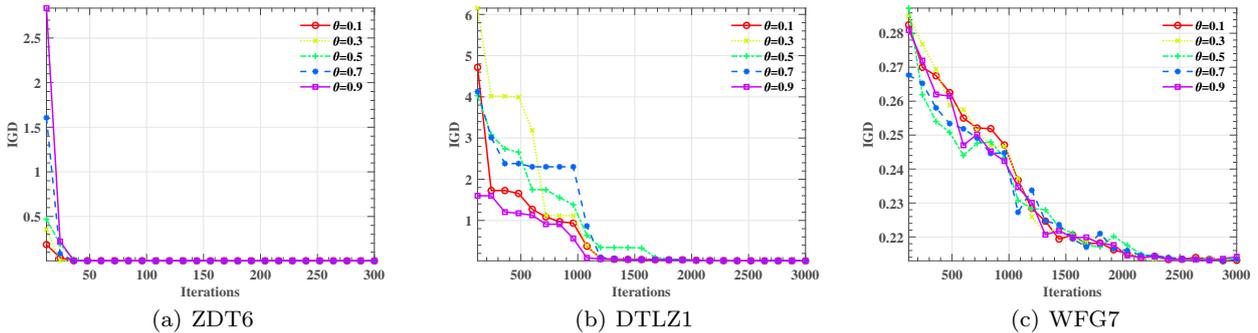

**Fig. 6** The mean IGD versus iterations obtained by the CMOMPA with different $\theta$ values on ZDT6, DTLZ1, WFG7 over 30 runs.

(a) ZDT6    (b) DTLZ1    (c) WFG7

SN can be deployed in a candidate location. It should be noted that in all experiments, the same deployment costs of locations are used. Unless otherwise stated, the parameter settings are listed in Table 9.

To evaluate the performance of the CMOMPA based optimal SN deployment method, we compared the coverage, connectivity, and deployment costs per day under different $K$-coverage and $C$-connectivity combinations. We assumed that each SN fails with a predefined probability as the working time increases. To be more realistic, the probability of SN failure also increases with the increase of working time, as shown in Table 10. Every 12 hours, we judge whether any surviving node is damaged according to the corresponding probability. We performed 10 independent repetitive experiments for each combination of $K$ and $C$ (See C3 and C4 in Eq. (25)).

Firstly, we investigate the effect of $K$ upon the coverage. Fig. 8 shows how the average coverage varies over network working time under different $C$ and $K$ combinations. The vertical dash lines indicate the number of days the full coverage of the monitoring are can be maintained. It can be seen that the time period for the network to maintain full coverage is improved with the increase of $K$. However, when $C = 3$, the improvement is not as significant as the scenarios of $C = 1$ and 2. This is because when $C = 3$, there are already enough SNs in the network, which satisfies the connectivity of the network and meanwhile improve the network coverage.

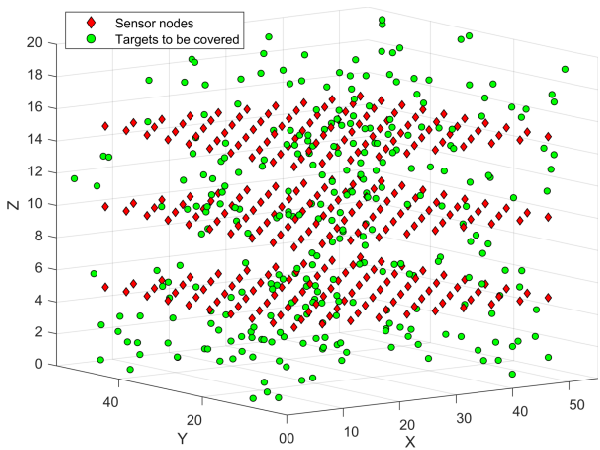

**Fig. 7** Deployment scenario of a WSN for simulations.

**Table 9** Simulation setups.

| Parameters | Values |
| --- | --- |
| Communication range, $R_c$ | 10 m |
| Measure of the uncertainty, $R_e$ | 2 m |
| Number of types of SNs | 3 |
| Deployment cost of SN of type 1, $C_{SN}(1)$ | 2 |
| Deployment cost of SN of type 2, $C_{SN}(2)$ | 5 |
| Deployment cost of SN of type 3, $C_{SN}(3)$ | 10 |
| Sensing range of SN of type 1, $R_s(1)$ | 5 m |
| Sensing range of SN of type 2, $R_s(2)$ | 10 m |
| Sensing range of SN of type 3, $R_s(3)$ | 15 m |
| Number of target points to be sensed | 300 |
| Parameter used in (7), $\lambda_1$ | 0.5 |
| Parameter used in (7), $\lambda_2$ | 1 |
| Threshold used in (8), $\delta$ | 0.8 |
| Relax threshold used in (25), $\varepsilon$ | 0.1 |
| Maximum number of iterations | 2000 |

Next, we further study how $C$ affects the network connectivity. Fig. 9 depicts the duration of network connectivity under various $C$ and $K$ combinations. We can read from it that, for a specific $K$, the duration of network connectivity increases with $C$. Similarly, for a specific $C$, the duration of network connectivity increases with $K$. The reason is straightforward since a greater $K$ or $C$ requires more SNs and thus enhance the network connectivity.

Finally, we investigate the daily cost under various $C$ and $K$ combinations. The daily cost of a WSN can be obtained by calculating the total deployment cost divided by its lifetime which is defined as follows:

**Definition 6** Lifetime of a WSN.
The lifetime of a WSN is defined as the duration (in days) that both the full coverage of the monitoring area (constraint C1 in Eq. (25)) and network connectivity (constraint C2 in Eq. (25)) are satisfied.

The results are shown in Fig. 10. It can be found that the lowest daily cost can be obtained in the case of $K = 1$ and $C = 2$, but not $K = 1$ and $C = 1$. The reason is that the low coverage and connectivity requirements lead to shorter lifetime of the WSN (see Fig. 8 and Fig. 9), which results in a higher daily cost. In scenarios with large $K$ and $C$, although the longer network lifetime can be achieved, the daily cost is considerably higher due to much higher total deployment cost.

We further compared with the PFs obtained by CMOMPA and four other top five algorithms listed in Table 7, namely CCMO, NSGA-III, MOMPA, DCNSGA-III. The results are shown in Fig. 11. We can see that the PF of CMOMPA dominates other algorithms, which indicates that the CMOMPA can find solutions with lower cost and higher reliability. It is worth noting that CCMO and DCNSGA-III failed to obtain PFs. Besides, NSGA-III only found one solution, rather than a Pareto surface. This is probably due to the huge dimension of constraints (there are thousands of constraints) in Eq. (25). The large-scale constraints limit their global search ability and make them fall into local infeasible regions.

The CMOMPA based deployment method can obtain a set of approximated Pareto optimal solutions, which helps the decision-makers choose the most appropriate one and balance the trade-off among deployment cost, sensing reliability and network reliability based on the actual application requirements. The followings are some examples.

*Example 1* An application needs to deploy a wireless sensor network to monitor the environment over a short period of time. Their budget is limited, but they need to achieve full coverage of the monitoring area and full connectivity of the network. In such a case, the solution with the maximum cost lower than the budget is the most proper one. It should be noted that with this solution, we can also ensure the maximum network lifetime, that is, the maximum $K$ and $C$ can be guaranteed under the given budget, since this solution is a non-dominated one.

*Example 2* Assume that the area where the WSN is deployed has high maintenance costs. In addition, we expect the network to work as long as possible. In this situation, the solution with the largest $K$ and $C$ is the expected one, because larger $K$ and $C$ prolong the network lifetime. Similarly, considering that the solution is non-dominated, the lowest cost can be also achieved with the same $K$ and $C$.

# 6 Conclusion

This paper balances the trade-off among the deployment cost, connectivity and coverage for heterogeneous

**Table 10** Probabilities of SN failure.

| Network lifetime (day) | 0-30 | 31-60 | 61-90 | 91-120 | 121-150 | 151-180 |
|---|---|---|---|---|---|---|
| Node death probability | 0.018 | 0.047 | 0.119 | 0.500 | 0.880 | 0.980 |

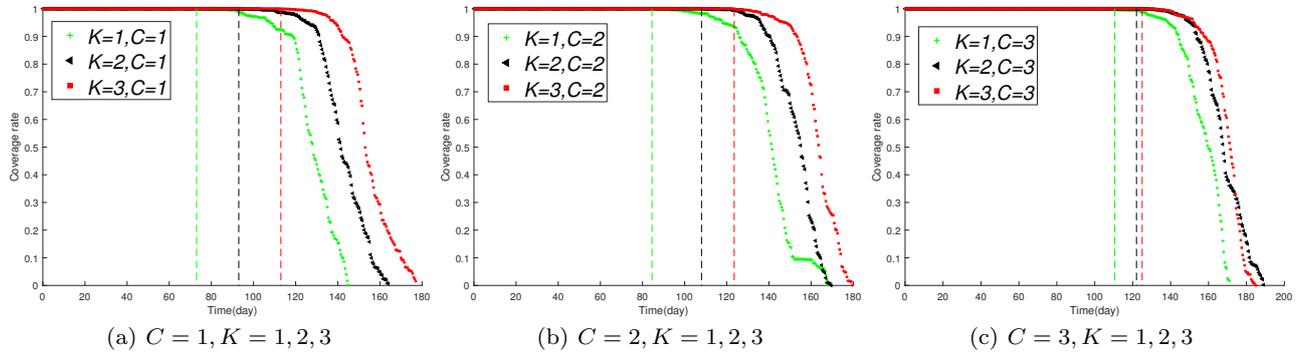

(a) $C = 1, K = 1, 2, 3$  (b) $C = 2, K = 1, 2, 3$  (c) $C = 3, K = 1, 2, 3$

**Fig. 8** Changes of coverage rates as the working time increases under various $C$ and $K$ combinations.

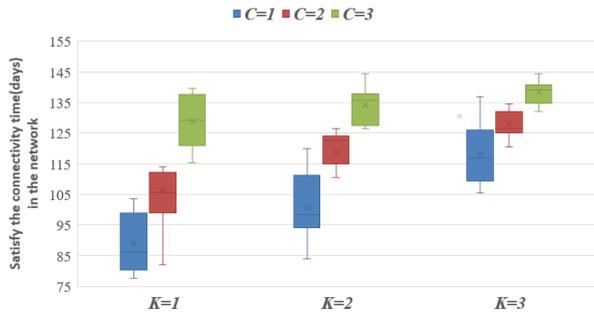

**Fig. 9** Durations of network connectivity under various $C$ and $K$ combinations.

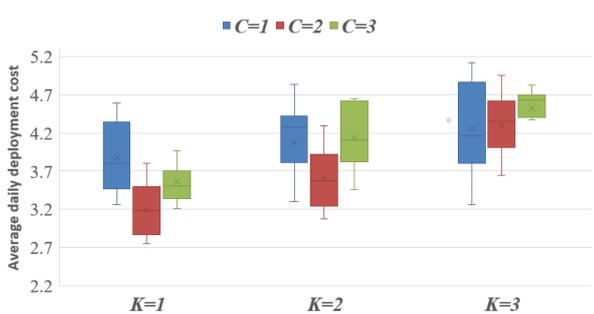

**Fig. 10** Daily costs under various $C$ and $K$ combinations.

WSNs under the constraints of specific coverage and network connectivity requirements. Different from the existing work, this paper fully considers the heterogeneity of SNs and 3-D environment, which is more suitable for practical WSNs deployments. In view that this is a very complicated multi-objective optimization problem, we propose a novel multi-objective swarm intelligence optimization algorithm, named CMOMPA, to solve it. CMOMPA is a multi-objective version of the MPA. It takes the reference points based method to determine elite individuals and leverages the Gaussian elitist perturbation strategy and competition mechanism-based learning strategy to produce offspring with stronger diversity and distribution. A large number of comparative computational studies over 20 benchmark functions show that CMOMPA has strong competitiveness. Based on CMOMPA, we propose the optimal SN deployment method for 3-D heterogeneous WSNs. Simulation results show that the proposed method can effectively obtain the approximated Pareto optimal solutions from which decision makers can choose the most appropriate solution and balance the trade-off among deployment cost, sensing reliability and network reliability based on the actual requirements of the application. It is worth noting that the CMOMPA proposed in this paper can also be easily used to handle other complex multi-objective optimization problems in engineering and scientific researches, only with necessary modifications on fitness functions, constraints and representations of individuals.

In future research, we will extend our work in the following aspects.

1) Firstly, we will consider the heterogeneity of communication ranges caused by various obstacles within the deployment areas. As known to all, the strengths of radio signals degrade when penetrating obstacles. So each SN may have a different communication range. Considering the heterogeneity of communication range can make our model more consistent with the actual situation.
2) Secondly, although CMOMPA proposed in this paper shows satisfactory performance in solving multi-modal multi-objective optimization problems, it is not effective enough for multi-objective optimization problems with irregular PFs or disconnected

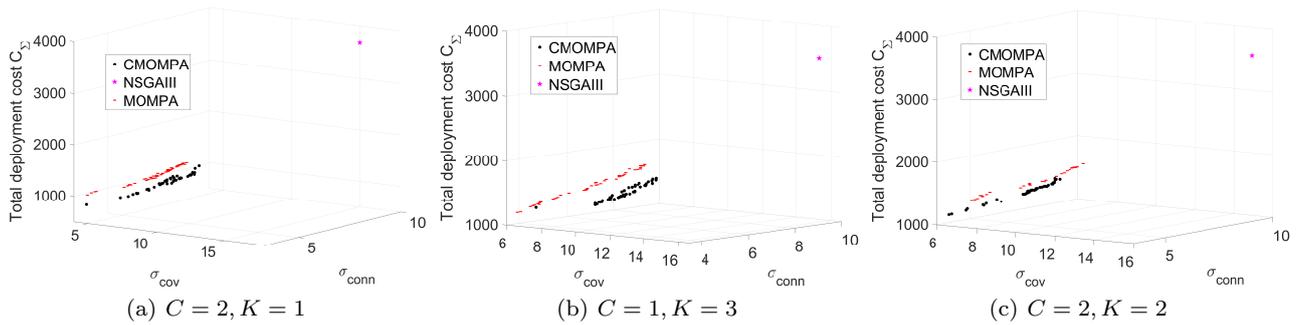

**Fig. 11** Comparisons on Pareto optimal fronts obtained by top five algorithms in Table 7. CCMO and DCNSGA-III failed to obtain PFs and NSGA-III only found one solution.

PFs. Therefore, improving CMOMPA so that it can deal with such multi-objective optimization problems is another focus of the follow-up research.

3) Thirdly, we are going to apply the proposed optimal deployment method to a real-world WSN applications, which is a large WSN to be deployed in a local large factory.

**Acknowledgements** This work was supported in part by the Natural Science Foundation of Zhejiang Province, China, under Grant LZ20F010008, and in part by the Xinmiao Talent Program of Zhejiang Province under Grant 2021R429025.